\documentclass[letterpaper,fleqn,twoside]{article}
\usepackage[colorlinks=true,breaklinks,bookmarks]{hyperref}
\usepackage{amsmath,amssymb}
\usepackage[dvips]{graphics,color,epsfig}
\renewcommand{\Im}{\mathop{\rm Im}}
\renewcommand{\Re}{\mathop{\rm Re}}

\begin{document}
\makeatletter
\renewcommand{\@evenhead}%
{\raisebox{0pt}[\headheight][0pt]{%
\vbox{\hbox to\textwidth{\thepage\hfil\strut {\small S.~A.~Myslivets et al.}}\hrule}}}
\renewcommand{\@oddhead}%
{\raisebox{0pt}[\headheight][0pt]{%
\vbox{\hbox to\textwidth{\strut {\small VUV at maximum coherence}\hfil\thepage
}\hrule}}} \makeatother

\title{\bf\small THEORY OF DEEP ULTRAVIOLET GENERATION AT MAXIMUM COHERENCE ASSISTED
BY STARK-CHIRPED TWO-PHOTON RESONANCE} \author{{\bf\small S.A.
Myslivets$^1$, A. K. Popov$^{1,2}$\thanks{E-mail: popov@ksc.krasn.ru;
URL: http://www.kirensky.ru/popov},
 V.V.Kimberg$^1$ and Thomas F. George$^2$%
 \thanks{Correspondence/Reprint request: Thomas F. George,Office of the Chancellor/Departments of Chemistry and
Physics \& Astronomy, University of Wisconsin-Stevens Point, Stevens
Point, WI 54481-3897, USA, E-mail: tgeorge@uwsp.edu;
URL: http://www.uwsp.edu/admin/chancell/tgeorge}}\\
\small $^1$Institute of Physics of Russian Academy of
Sciences, 660036 Krasnoyarsk, Russia\\
\small $^2$Office of the Chancellor~/~Departments of Chemistry and
Physics~\&~Astronomy,\\  \small University of Wisconsin-Stevens
Point, Stevens Point, WI 54481-3897, USA\\
\\
\small A chapter in {\it Modern Topics in Chemical Physics}, \\
\small ed. by T. F. George, X. Sun and G. P. Zhang,\\ \small
(Research Signpost, Trivandrum, India, 2002)\\
\\Dated: October 6, 2001}
\date{}
\maketitle

{\it A scheme is analyzed for efficient generation of vacuum ultraviolet radiation
through four-wave mixing processes assisted by the technique of Stark-chirped rapid
adiabatic passage. These opportunities are associated with pulse excitation of
ladder-type short-wavelength two-photon atomic or molecular transitions so that
relaxation processes can be neglected. In this three-laser technique, a delayed-pulse
of strong off-resonant infrared radiation sweeps the laser-induced Stark-shift of a
two-photon transition in a such way that facilitates robust maximum two-photon
coherence induced by the first ultraviolet laser. A judiciously delayed third pulse
scatters at this coherence and generates short-wavelength radiation.  A theoretical
analysis of these problems based on the density matrix is performed. A numerical model
is developed to carry out simulations of a typical experiment. The results illustrate
a behavior of populations, coherence and generated radiation along the medium as well
as opportunities of efficient generation of deep (vacuum) ultraviolet radiation.}


\section{Introduction}
A technique of nonlinear-optical frequency conversion in gaseous
media provide a well-established powerful tool for the generation of
tunable coherent short-wavelength vacuum-ultraviolet (VUV) radiation
\cite{Wal82}. Many applications have been found for these light
sources especially in spectroscopy. However, due to the relatively
small off-resonance non-linear susceptibilities and severe problems
associated with absorption and dispersion in resonance atomic or
molecular gases conversion efficiencies obtained with this technique
is rather small, typically in the range of 10$^{-6}$-10$^{-4}$.  (For
a review, see e.g. \cite{{Ark87,Hel81}}). Negative accompanying
resonance effects can be reduced by manipulating the
coherently-driven medium through {\it nonlinear interference effects}
at quantum transitions (see e.g. the review in
\cite{Rau96,AKP96,AKP98}) in order to minimize resonant absorption
and maximize non-linear susceptibilities. One of the manifestations
of destructive interference at quantum transitions is {\it
electromagnetically induced transparency} (EIT) \cite{Har90,Mar98}
and {\it coherent population trapping} \cite{Ari96}, which lead to
the cancellation of absorption for some of the coupled fields and may
improve phase matching. By the appropriate choice of detuning and
intensities, VUV radiation that is generated in a frequency
conversion process has been produced  with conversion efficiencies up
to 10$^{-2}$ \cite{Zhang93,Dor00,Dor00a}. In strongly-driven media,
one cannot think about nonlinear-optical response in terms of
susceptibilities in their original sense. While EIT may lead to
increased conversion efficiencies, the use of large coupling fields
to produce transparency in a Doppler-broadened medium is in conflict
with the requirements for maximizing the nonlinear susceptibility.
Despite the constructive interference, the large Autler-Townes
splitting results in much reduced nonlinear susceptibilities compared
to resonant values. Furthermore, when tunability is required,
resonances with the generated field cannot be used  to provide an
enhancement in the nonlinear optical response of the medium.

In a two-level atom, optical polarization depends on the coherence
between the ground and excited states, and reaches a maximum of 1/2
for an equal amplitude coherent superposition of ground and excited
states. Therefore, in a medium comprised of a high density of
dark-state atoms,  a relatively-large nonlinear polarization can be
produced while absorption for the driving resonant fields is
decreased. The system may then be regarded as an oscillator driven
with maximum amplitude. If another electromagnetic field is
introduced, the beat frequency produced by interaction with the
oscillator may be generated with high conversion efficiency, which in
principle can approach unity. In this system the nonlinear
polarization responsible for generation may become as large as the
terms responsible for absorption. Thus, nonlinear interference
processes in media with maximum coherence can substantially improve
frequency conversion between the coupled fields. Frequency-mixing
processes at maximum coherence have been studied theoretically
\cite{Hem95,Har96,Har97,Luk98,Man98,Baev00} as well as demonstrated
experimentally \cite{Hem95,Har96,Hak97,Har99} both with adiabatic
laser pulses and for steady-state conditions. In the latter case, the
lowest relaxation rate and therefore maximum coherence is achievable
at the Raman transitions associated with the ground electronic state.
Consequently, in four-wave mixing experiments, maximum coherence has
been established in a $\lambda$-type coupling scheme, similar to that
employed for coherent anti-Stokes Raman scattering (CARS).
$\lambda$-type coupling gives limits to the shift of the generated
radiation to the short-wavelength ranges compared to the ladder-type
schemes. Besides that, laser-induced Stark shifts, which are
intrinsic to strong coupling, may perturb the adiabatic population
dynamics and prohibit the preparation of  maximum coherence
\cite{Boh98,Guer,Boh01}.

When the pulse duration is much shorter than the coherence relaxation
rates in a quantum system, coherent quantum control of populations
based on {\it Stark-chirped rapid adiabatic passage} (SCRAP) can be
performed that does not require Raman type ground-state coherence.
Physical principles of this promising technique and
proof-of-principle experiments are described in a recent publication
\cite{Ric00}. In SCRAP, two-photon (in the general case,
multi-photon) detuning of the driving field is controlled through a
Stark shift of the upper level by an auxiliary strong laser pulse
$E_{St}(\omega_{St})$, which is time-shifted with respect to the pump
laser. The process is closely related to rapid adiabatic passage
controlled by a frequency chirp in the pump laser \cite{Loy76,Loy78}.

This chapter investigates potentials and features of deep (vacuum
ultraviolet) generation at maximum coherence based on the SCRAP technique
\cite{Ric00}. However, instead of maximum population transfer from the
ground to an excited state under the driving field $E_1(\omega_1)$, the
emphasis is placed on generating persistent maximum coherence at the
frequency $2\omega_1$. This is related to robust transfer of half of the
population from the ground to the two-photon-resonant upper state. This
process is substantially effected by a time-dependent Stark shift of the
two-photon resonance induced by both control (auxiliary) and fundamental
(pump) time-shifted pulses. A third probe field $E_2(\omega_2)$ scatters at
oscillations at $2\omega_1$ and generates difference- and sum-frequency deep
UV radiation at $\omega_{\mp}=2\omega_1\mp\omega_2$.

We employ a numerical simulation in order to illustrate the dynamics of the
system, dependent on the combination of laser parameters, as well as
propagation effects of the coupled driving (fundamental), weak convertible
and generated electromagnetic waves in an optically-dense medium. The
spacial behaviors of both sum-frequency and difference-frequency processes
are considered, while the fundamental radiation and consequently  medium
properties may vary along its length.

\section{Principal equations}
\begin{figure}
\centering\includegraphics*[width=40mm]{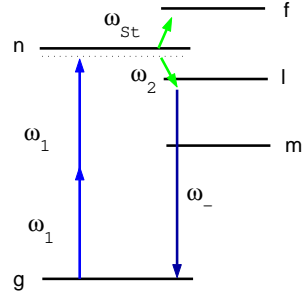} \caption{ Energy
levels and coupled fields. }\label{4lsh}
\end{figure}

A partial energy-level scheme relevant to the above described four-wave-mixing process
$\omega_{\mp}=2\omega_1\mp\omega_2$ controlled by a strong auxiliary infrared
off-resonant field $E_{St}$ is depicted in Fig.~\ref{4lsh}.
Levels $m$ and $l$ give the major contribution to the corresponding two-photon quantum
pathways, and level $f$ to Stark shift. When analyzing sum-frequency process at
$\omega_+$, level $l$ is placed above  $n$ with the same three-photon detuning.  All
the fields are assumed pulsed, where the duration of each pulse is much shorter than
the shortest relaxation time in the quantum system. Through a time-dependent
laser-induced Stark shift of
 level $n$, i.e., SCRAP, the field $E_{St}$ allows the control of
two-photon excitation by the field $E_1$. By this process, the entire
population of the ground level can be moved to the upper state $n$ by the
end of the pulse $E_1$. Alternatively, maximum coherence at the two-photon
transition $gn$ can be achieved by making the populations of states $g$ and
$n$ equal by the end of the pulse $E_1$. Furthermore, with a judicious delay
of the pulse $E_2$, one can achieve maximum nonlinear polarization at the
frequency $\omega_{\mp}$ and therefore high conversion efficiency of the
visible radiation to the VUV range.

\subsection{Equations for interacting electromagnetic fields}
The propagation of optical waves in a non-conducting medium is described by
the wave equation
\begin{equation}\label{urmax}
\nabla^2 \mathcal E-\frac1{c^2}\frac{\partial^2 \mathcal E}{\partial t^2}=
\frac{4\pi}{c^2}\frac{\partial^2\mathcal P}{\partial t^2},
\end{equation}
where $\mathcal P$ is the electrical polarization,and we assume the
nonlinear medium to be isotropic and all the waves to be identically
polarized and to propagate along the $z$-axis:
\begin{eqnarray}\label{field}
 \mathcal E_j(z,t)=\Re \{E_j(z,t)\exp[i(\omega_jt-k_iz)]\},\nonumber\\
  \mathcal P_j(z,t)=\Re \{P_j(z,t)\exp[i(\omega_jt-k'_iz)]\}.
\end{eqnarray}
Here $k_j$ is the modulus of a wave vector at frequency $\omega_j$, $k'_j$
is the wave vector of nonlinear polarization at the same frequency, and
$E_j(z,t)$ and $P_j(z,t)$ are slowly-varying envelopes.

Assuming a uniform field distribution over the cross section and taking into
account Eqs. (\ref{field}) and $k\approx\omega/c$, one can write Eq.
(\ref{urmax}) in the approximation of slowly-varying amplitudes as
\begin{equation}\label{urmax1}
2ik\left(\frac{\partial E}{\partial z}+\frac{1}{c}\frac{\partial E}{\partial
t}\right)=-4\pi k^2 P,
\end{equation}
where $P=P^{L}+{P}^{(FWM)}e^{-i\Delta kz}$, and $\Delta k$ is the phase
mismatch of the interacting waves. We restrict ourselves to the major
process under consideration and  assume phase matching to be fulfilled so
that $\Delta k=0$. Then by setting $z'=z$, $t'=t-z/c$ and using
\begin{eqnarray*}
\frac{\partial E}{\partial z}&=&\frac{\partial E}{\partial z'}\frac{\partial
z'}{\partial z} + \frac{\partial E}{\partial t'}\frac{\partial t'}{\partial
z}= \frac{\partial E}{\partial z'}-\frac1c\frac{\partial E}{\partial
t'},\\
\frac{\partial E}{\partial t}&=&\frac{\partial E}{\partial t'}\frac{\partial
t'}{\partial t} + \frac{\partial E}{\partial z'}\frac{\partial z'}{\partial
t}=\frac{\partial E}{\partial t'},\nonumber
\end{eqnarray*}
we can express Eq.~(\ref{urmax1}) in the form
\begin{equation}\label{urmax2}
{\partial E}/{\partial z'}=2\pi i k P.
\end{equation}
The polarization $P$ can be calculated with the aid of the density matrix
$\rho_{ij}$ as \[P=N\rho_{ij}d_{ji}+\hbox{c.c.},\] where N is number density
of the resonance atoms,  $d_{ji}$ are electro-dipole transition matrix
elements, and $\rho_{ij}$ are density matrix components oscillating at the
corresponding frequencies. The problem then reduces to calculating the
off-diagonal elements of the density matrix.
\subsection{Density matrix equations}
In the case where the pulse durations of all the interacting fields are much
less than any relaxation time in the system, all the relaxation terms may be
omitted, and the density matrix equations take the form
\begin{eqnarray}\label{rho}
\dot\rho_{mn}&=&-i\left[G_{mn}\rho_n+G_{mg}\rho_{gn}\right],\nonumber\\
\dot\rho_{ln}&=&-i\left[G_{ln}\rho_n+G_{lg}\rho_{gn}\right],\nonumber\\
\dot\rho_{gl}&=&i\left[\rho_{gn}G_{nl}+\rho_{g}G_{gl}\right],\nonumber\\
\dot\rho_{gm}&=& i\left[\rho_{gn}G_{nm}+G_{gm}\rho_{g}\right]\nonumber\\
\dot\rho_{gn}&=&-i[G_{gm}\rho_{mn}-G_{mn}\rho_{gm}\nonumber\\
&+& G_{gl}\rho_{ln}- G_{ln}\rho_{gl}-G_{fn}\rho_{gf}]\nonumber\\
\dot\rho_n&=&2\Im\left[G_{nm}\rho_{mn}+G_{nl}\rho_{ln}+G_{nf}\rho_{fn}\right],\nonumber\\
\dot\rho_{nf}&=&iG_{nf}\rho_n,\, \dot\rho_{gf}=i\rho_{gn}G_{nf},\, \rho_g=1-\rho_n.
\end{eqnarray}
Here $\dot\rho_{ij}=d\rho_{ij}/dt$ (hereafter we omit the prime on $z$ and
$t$, assuming a transformation to a new coordinate system),
$G_{ij}(t)=-d_{ij}E_k(t)/2\hbar$ are the coupling Rabi frequencies, and
$d_{ij}$ are electric dipole transition momenta. In the coupled equations
(\ref{rho}), we assume that initially only the lower level $g$ is populated
and that the populations of levels $m$, $l$ and $f$ are negligibly small
during the entire process. We also neglect photo-ionization. Further, we
assume two-photon quasi-resonance
$\Omega_{gn}=(2\omega_1-\omega_{gn})\ll\omega_{ij}$ and that the conditions
for one-photon detunings
\begin{equation}\label{*}
\Omega_{ij}=(\omega_k-\omega_{ij})\gg G_{ij}\gg \tau_i^{-1}
\end{equation}
are fulfilled, where $\tau_i$ is the minimum pulse length.

We look for the solution of the coupled equations in the form
\begin{eqnarray*}
\rho_{ij}(t)&=&r_{ij}(t)\exp\left\{-i\Omega_{ij}t\right\},\nonumber\\
G_{ij}(t)&=&g_{k}(t)\exp\left\{-i\Omega_{ij}t\right\},\nonumber\\
 G_{mn}(t)&=&ag_{1}(t)\exp\left\{-i(\Omega_{mn}t\right\}, \nonumber
\end{eqnarray*}
where $r(t)$ and $g(t)$ are slowly-varying envelopes, and $a=d_{mn}/d_{gm}$.
Then using (\ref{*}) and $|\dot r_{ij}|\ll|\Omega_{ij}r_{ij}|$ for all the
off-diagonal elements except $r_{gn}$,  we can rewrite the equations
(\ref{rho}) as
\begin{eqnarray}\label{r}
r_{mn}&=&\left[ag_1 r_n+g_1^* r_{gn}\right]/\Omega_{mn},\nonumber\\
r_{ln}&=&\left[g_2 r_n+g_-^* r_{gn}\right]/\Omega_{ln},\nonumber\\
r_{gl}&=&-\left[r_{gn}g_2^*+r_{g}g_-\right]/\Omega_{gl},\nonumber\\
r_{gm}&=&-\left[a^*g_1^* r_{gn}+g_1 r_{g}\right]/\Omega_{gm},\nonumber\\
\dot r_{gn}&=&-i[g_1 r_{mn}-ag_1 r_{gm}+g_- r_{ln}\nonumber\\
&-&g_2 r_{gl}-g_{St}^* r_{gf}-\Omega_{gn}r_{gn}],\nonumber\\
\dot r_n&=&2\Im\left[a^*g_1^* r_{mn}+g_2^* r_{ln}+g_{St}
r_{fn}\right],\nonumber\\
r_{nf}&=&-g_{St}r_n/\Omega_{nf},\, r_{gf}=-r_{gn}g_{St}/\Omega_{gf},\nonumber\\
r_g&=&1-r_n.
\end{eqnarray}
Here $\Omega_{nf}= \omega_{St}-\omega_{nf}$, $\Omega_{gl}=
2\omega_{1}-\omega_2-\omega_{gl}$, $\Omega_{gf}= 2\omega_{1}+ \omega_{St}
-\omega_{gf}$, etc. After simple algebra, Eqs. (\ref{r}) reduce to
\begin{align}\label{dif}
\dot
r_n&=2\Im\left[\left(a^*\frac{{g^*_1}^2}{\Omega_{mn}}+\frac{{g_2^*g_-^*}}
{\Omega_{ln}}\right)r_{gn}\right],\nonumber\\
\dot r_{gn}&=-i[\left(\Omega_{St}-\Omega_{gn}\right)r_{gn} +g_{gn}r_n\nonumber\\
&+\frac{ag_{1}^2}{\Omega_{gm}}+\frac{g_2g_{-}}{\Omega_{gl}}],
\end{align}
where
\begin{align}\label{RSt}
g_{gn}&=\left(\frac{ag_{1}^2}{\Omega_{mn}}+\frac{g_2g_{-}}{\Omega_{ln}}
-\frac{ag_1^2}{\Omega_{gm}}-\frac{g_2g_{-}}{\Omega_{gl}}\right),\nonumber\\
 \Omega_{St}&=\frac{|g_-|^2}{\Omega_{ln}}
+\frac{|g_{St}|^2}{\Omega_{gf}}+\frac{|g_{1}|^2}{\Omega_{mn}}
+\frac{a^2|g_{1}|^2}{\Omega_{gm}}+\frac{|g_2|^2}{\Omega_{gl}}.
\end{align}
\subsubsection{Generalized two-level scheme}
In the more general case of multi-level system a sum over all intermediate
states $m$ and $l$ must be taken. In  view of that, we introduce the values
\begin{align}\label{RabSt}
r_1&=- 2\frac{ag_1^2}{\Omega_{gm}}=\gamma_1E_1^2,\, s=\frac{|g_{St}|^2}
{\Omega_{nf}}=\beta_S|E_{St}|^2,\nonumber\\
s_1&=\left(\frac{a^2}{\Omega_{gm}} +
\frac{1}{\Omega_{mn}}\right)|g_1|^2= \beta|r_1|=\beta_1|E_1|^2,\nonumber\\
r_2&=-2\frac{g_2g_-}{\Omega_{gl}}=\gamma_2E_2E_{-},\,
s_2=\frac{|g_{2}|^2}{\Omega_{gl}}=\beta_2|E_{2}|^2,  \nonumber\\
s_-&= \frac{|g_{-}|^2}{\Omega_{ln}}=\beta_-|E_{-}|^2,
\end{align}
where $\gamma_i$ and $\beta_i$ are polarizabilities depending on specific
atoms and contributing quantum transitions. In the vicinity of a two-photon
resonance $\Omega_{mn}\approx -\Omega_{gm}$ and $\Omega_{ln}\approx
-\Omega_{gl}$. Then Eqs. (\ref{dif}) take the form:
\begin{align}\label{2l}
\dot r_n&=\Im\left[\left(r_1^*+r_2^*\right)r_{gn}\right],\nonumber\\
\dot r_{gn}&=-i\left(\Omega_{St}-\Omega_{gn}\right)r_{gn}\nonumber\\
&-i\left(r_1+r_2\right)\left(r_{n}-1/2\right),
\end{align}
where
\begin{eqnarray}\label{SumSt}
\Omega_{St}=s + s_1 + s_2+ s_-.
\end{eqnarray}
It is seen from Eqs. (\ref{dif}) that $\Omega_{St}$ is the laser-induced
shift of the two-photon resonance.
\subsection{Evolution of Rabi frequencies along the medium}
We assume that {\it one-photon} Rabi frequencies of all the fields vary in
time as
\begin{equation}\label{g}
g_i(t)=g_{i0}\exp\{-(t-\Delta\tau_i)^2/2\tau_i^2\}. \end{equation} Here
$2\tau_i$ is the width of the $i$-th pulse (at $e^{-1}$ of the power
maximum), and $\Delta\tau_i$ is the time delay of the $i$-th pulse relative
to $t=0$. The pulse widths are assumed to be much shorter than the
relaxation rates of the system. We also assume that the spectral width of
the pulse $E_1$ is much greater than the Doppler width of the transition
$gn$, and are replaced Eq. (\ref{urmax2}) by the equations for the
corresponding Rabi frequencies,
\begin{eqnarray*}
{d\,g_1}/{d\,z}&=&-i\pi\hbar^{-1} k_1 |d_{gm}|^2N
(r_{gm}+a^* r_{mn})\\
&=&-i\alpha_{1}(r_{gm}+a^* r_{mn}),\\
{d\,g_2}/{d\,z}&=&-i\alpha_2 r_{ln}, \quad
{d\,g_{-}}/{d\,z}=-i\alpha_{-}r_{gl},
\end{eqnarray*}
where $\alpha_{1,2,-}$ are resonant frequency-integrated absorption indices
at the corresponding transitions for broadband probe radiations, with all
other fields turned off.  Then by scaling the medium length  to the
absorption length $\alpha_{-}^{-1}$, which is a characteristic parameter for
the given medium that can be measured independently, we obtain
\begin{eqnarray}\label{urmaxg}
{d\,g_1}/{d\xi}&=&-iK_{1}(r_{gm}+a^* r_{mn}),\nonumber\\
{d\,g_2}/{d\xi}&=&-iK_2r_{ln},\nonumber \\
{d\,g_-}/{d\xi}&=&-ir_{gl},
\end{eqnarray}
where
\begin{eqnarray}\label{K}
K_{1}&=&(k_1|d_{gm}|^2)/(k_-|d_{gl}|^2)\nonumber\\
&=&
(k_1\omega_{gl}|f_{gm}|)/(k_{-}\omega_{gm}|f_{gl}|),\nonumber\\
K_2&=&(k_2|d_{ln}|^2)/(k_{-}|d_{gl}|^2)\nonumber\\
&=&(k_2\omega_{gl}|f_{ln}|)/(k_{-}\omega_{ln}|f_{gl}|),\nonumber\\
\xi&=&\alpha_{-}z.
\end{eqnarray}

In the above consideration we ignored Doppler effects, which is possible,
for example, in atomic beams.  In the case of warm gas, the solution of the
equations must be averaged over a Maxwell velocity distribution of atoms.
However, Doppler effects do not contribute much if the equivalent spectral
width of the pulses or/and all detunings are greater or comparable with the
corresponding resonance Doppler HWHM.
\section{Numerical simulation of SCRAP and maximum coherence in a generalized
two-level scheme}
\begin{figure}[!h]
\begin{center}
\includegraphics[width=.3\columnwidth]{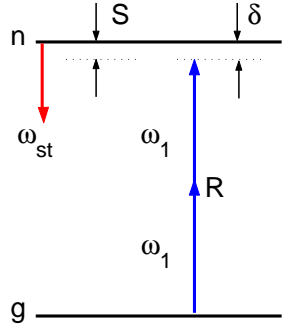}
\caption{\label{2lsh} Generalized two-level scheme and radiative couplings.
}
\end{center}
\end{figure}

For further numerical simulations we introduce the parameters $\delta
=\Omega_{gn}\tau_1$, $S=s_0\tau_1$, $R=r_{10}\tau_1$,  $T = t/\tau_1$ and
$\delta\tau=\Delta\tau/\tau_1$, which are the detuning and maximum dynamic
shift of the resonance scaled to the spectral width of the pulse $E_1$ (Eqs.
(\ref{g})), number of Rabi oscillations during the pulse duration $\tau_1$,
and time and pulse shifts scaled to the duration of this pulse. In this
subsection we analyze the dynamics of population transfer and induced
coherence, assuming $g_{2}=0$ and the dynamic self-shift of the resonance to
be negligible ($\beta=0$). Then the equations (\ref{2l}) describe
laser-induced oscillations in a two-level system controlled by the auxiliary
Stark field. The corresponding level scheme and radiative coupling are shown
in Fig. \ref{2lsh}.
%
%
%
\subsection{Oscillations induced by a rectangular pulse}
First, we simulate the well-known dynamics induced by pulses of rectangular
shape and duration $\tau$ at  $\Omega_{gn}=\Omega_{St}$ ($\pi/2$, $\pi$ and
$2\pi$ pulses). Then the  equations (\ref{2l}) take the form:
\begin{equation}\label{Rabi}
\dot r_{gn}=-ir_{1}(r_n-1/2),\quad \ddot r_n= -|r_1|^2(r_n-1/2).
\end{equation}
Equation (\ref{Rabi}) describes induced oscillations with the two-photon
Rabi frequency $|r_{1}|$. The coherence $r_{gn}$ reaches its maximum ($\dot
r_{gn}=0$) at $r_n=1/2$. Then $r_g=1-r_n=1/2$ too, and $r_{gn}$, which is
product of the corresponding probability amplitudes, reaches its maximum
$|r_{gn}|=1/2$ as well.

\begin{figure}[!h]
\vspace{-3mm}\begin{center}
\includegraphics[width=.6\columnwidth]{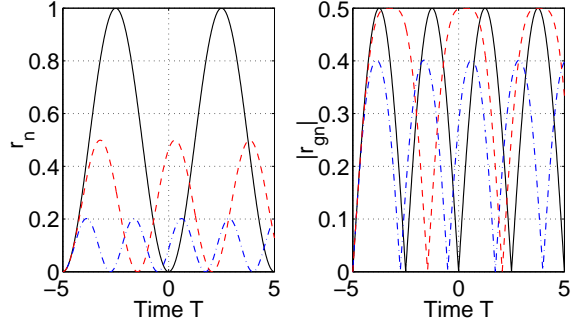}
\caption{\label{rabi} Dependence of the population $r_n$ and coherence
$|r_{gn}|$ on time $T$ in a generalized two-level scheme with excitation by
a {\it rectangular pulse} at various deviations from resonance
$\Omega_{gn}$. Time is scaled to the arbitrary value $\tau$ which is much
less then any relaxation rate in the system. The coupling Rabi frequency
$R=r_1\tau=2\pi/5$, and $\delta =\Omega_{gn}\tau$ are scaled to $\tau^{-1}$.
Solid line -- $\delta=0$, dashed -- $\delta = 1.259$, dash-dotted -- $\delta
= 2.5$. }
\end{center}
\end{figure}

At $\Omega=\Omega_{gn}-\Omega_{St}=0$ and $t=\pi/2|r_1|$, half of the
population of the ground state transfers to the excited state, and the
amplitude of the polarization oscillations $|r_{gn}|$ reaches its maximum.
At $t=\pi/|r_1|$ the entire population of the ground state is transferred to
the excited state. At $t=3\pi/2|r_1|$ the coherence again reaches its
maximum, and at $t=2\pi/|r_1|$ the system returns to its initial state. Such
oscillations are illustrated in Fig. \ref{rabi}. Deviation from resonance
leads to increased frequency and decreased amplitude of the oscillations for
the level populations (Fig. \ref{rabi} (left)), and  the instant when
maximum coherence is reached will vary. However, the maximum value of
coherence, which is 0.5, is not achievable if the population of the upper
level does not reach 0.5. (Fig. \ref{rabi} (right)).
\vspace{0mm}
\vspace{-2mm}
\subsection{Dynamics of generalized two-level system driven by Gaussian
pulses}
In further analysis we shall account for the differences which appear for
the pulses of {\it Gaussian shape} (Eq. (\ref{g})). This leads to variation
in time of the {\it two-photon} Rabi frequency and Stark shifts as:
\begin{eqnarray}
r_i(t)=r_{i0}\exp\left[-(t-\Delta\tau_{i})^2/\tau_{i}^2\right],\nonumber\\
s(t)=s_{0}\exp\left[-(t-\Delta\tau_{St})^2/\tau_{St}^2\right],\nonumber\\
s_{i}(t)=s_{i0}\exp\left[-(t-\Delta\tau_{i})^2/\tau_{i}^2\right].
\label{Stt}
\end{eqnarray}
Hereafter, we will scale a static deviation from the two-photon resonance
$\Omega_{gn}$, amplitudes of the dynamic resonance shift $s_0$ and Rabi
frequency $r_{10}$ to the spectral width $\tau_1^{-1}$, and delay of the
Stark pulse $\Delta\tau$ and time $t$ to the duration of the excitation
pulse $\tau_1$: $\delta =\Omega_{gn}\tau_1$, $S=s_0\tau_1$,
$R=r_{10}\tau_1$, $\delta\tau=\Delta\tau/\tau_1$, $T = t/\tau_1$.
\subsubsection{Persistent maximum coherence created by Gaussian
pulses at negligible dynamic self-shift of the resonance}
\begin{figure}
\begin{center}
\includegraphics[width=.6\columnwidth]{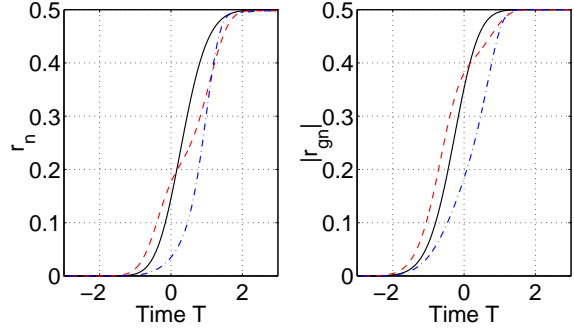}
\vspace{2mm} \caption{\label{2lst} Population $r_n$ and persistent maximum
coherence $|r_{gn}|$ controlled by the dynamic Stark shift at various static
deviations from the two-photon-resonance and moderate two-photon Rabi
frequencies. The Stark shift produced by the two-photon excitation is
ignored. The duration of the control Gaussian Stark-shift pulse is
$\tau_{St}/\tau_1=1.6$. Solid plot --- $\delta=0$, $S=0$, $R=0.886$;  dashed
--- $\delta=5$, $S=7.4$, $R=3.18$, $\delta\tau=1.8$; dash-dotted ---
$\delta=20$, $S=23$, $R=3.48$, $\delta\tau=1.34$.}
\end{center}
\end{figure}
In the case under consideration in this sub-subsection ($\beta=0$), the
principal contribution to $\Omega_{St}$ is determined by $E_{St}$, such that
the entire Stark shift depends on time as
\begin{eqnarray}
\Omega_{St}(t)&=&s_{0}\exp\left[-(t-\Delta\tau_{St})^2/\tau_{St}^2\right].
\label{2lSt}
\end{eqnarray}
We choose the intensity and  frequency detuning of the Stark field such that
the Stark shift $\Omega_{St}$ can compensate for the two-photon detuning
$\Omega_{gn}$.  In order to provide a dynamic two-photon resonance, the
requirement $|s_{0}|>|\Omega_{gn}|$ must be fulfilled. Then the resonance
condition is met at two instants of time:
\begin{equation}\label{time}
 t_{1,2}=\Delta\tau_{St}\mp\tau_{St}\sqrt{\ln({s_{0}}/{\Omega_{gn})}}.
\end{equation}
By a proper choice of the field parameters, one can provide first passage of
the two-photon resonance at such an instant that the value of the two-photon
Rabi frequency ensures a $\pi/2$ process over the period of passage through
the resonance. For example, if the first instant of the crossing of the
resonance is $t_1=0$ and the second one is $t_2=2\Delta\tau_{St}$, then the
requirement is:
\begin{equation}\label{det}
\Omega_{gn}=s_{0}\exp(-\Delta\tau_{St}^2/\tau_{St}^2)
\end{equation}

A numerical analysis of corresponding feasibilities of creation of
persistent maximum coherence is presented in Fig. \ref{2lst}. Figure
\ref{2lst} (solid) illustrates the maximum coherence generated by the
resonance $\pi/2$ pulse with the Stark field turned off. Figures
\ref{2lst} (dash and dash-dotted) show such opportunities realized
with the aid of the Stark field at various deviations of excitation
from the two-photon resonance. The Stark shift at the front of the
pulse is adjusted to compensate for the static deviation at the
instant when the excitation is around its maximum values, while the
Rabi frequency is about the magnitude required to transfer half of
the population of the ground state to the excited one during the
period of passage through the resonance. Due to the appropriately
selected pulse shift, the Rabi frequency is too small for the
population to return to the ground state while passing through the
resonance at the rear of the Stark pulse.
\begin{figure}
\begin{center}
\includegraphics[width=.5\columnwidth]{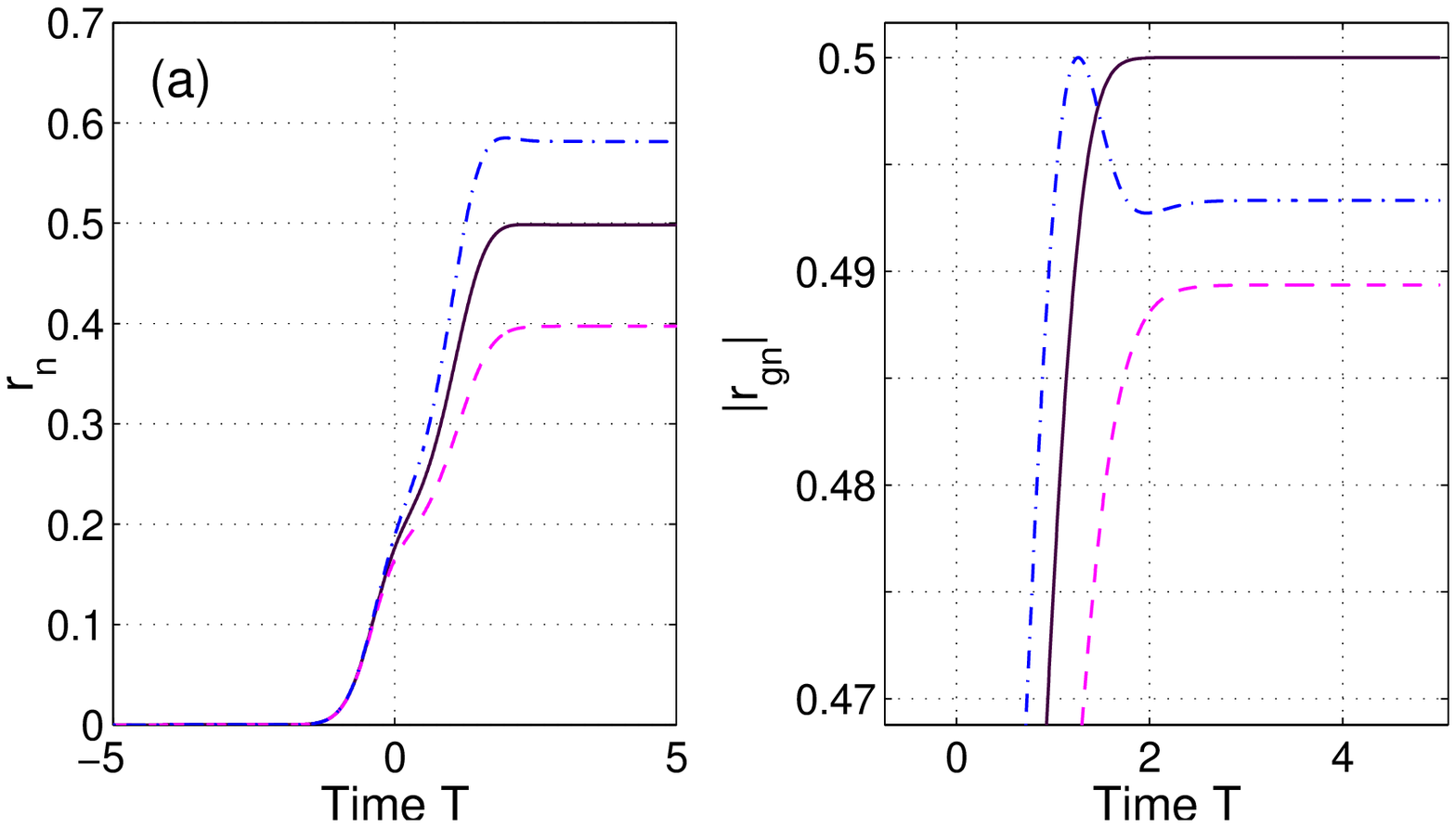}\\
\includegraphics[width=.5\columnwidth]{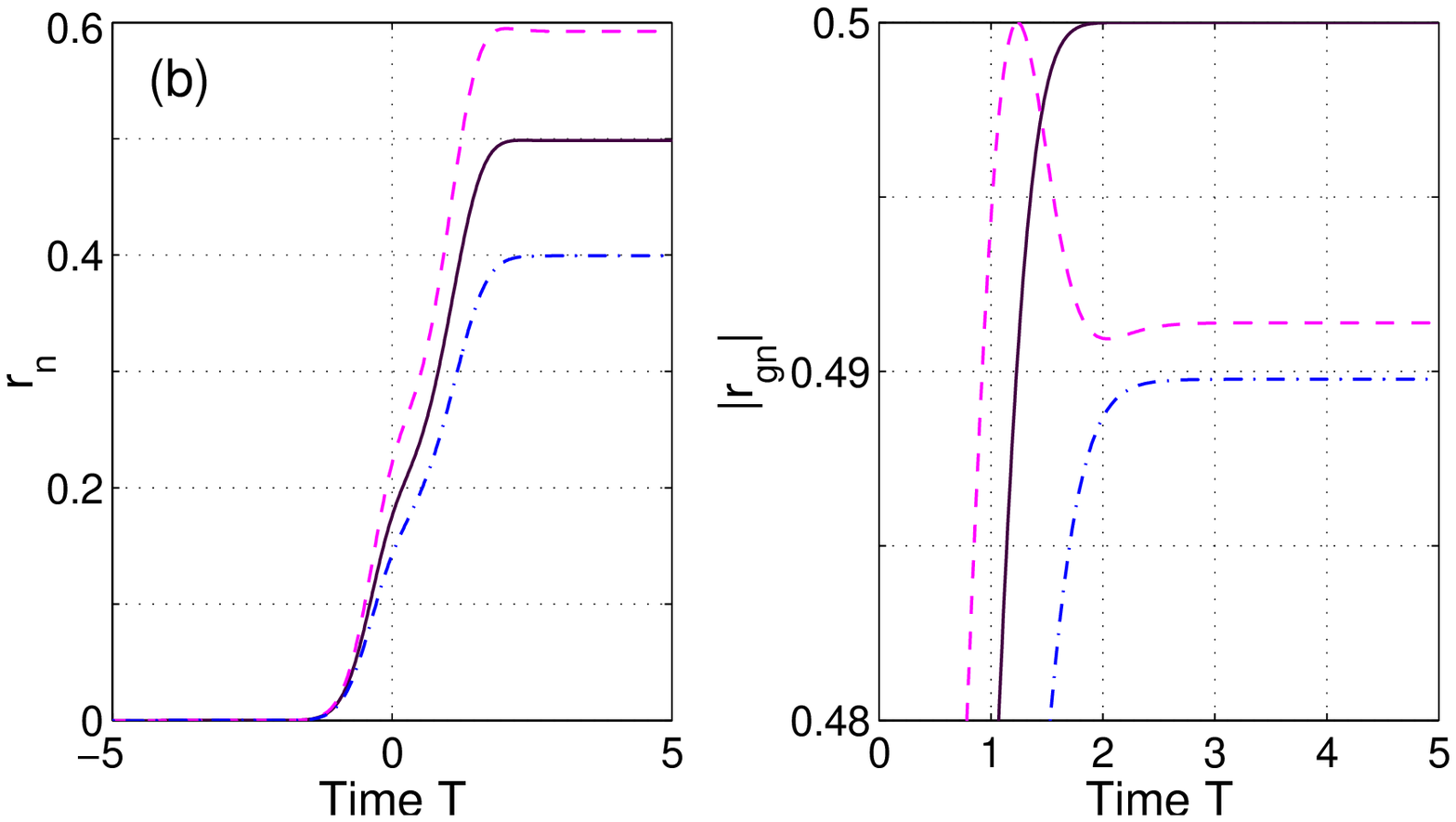}\\
\includegraphics[width=.5\columnwidth]{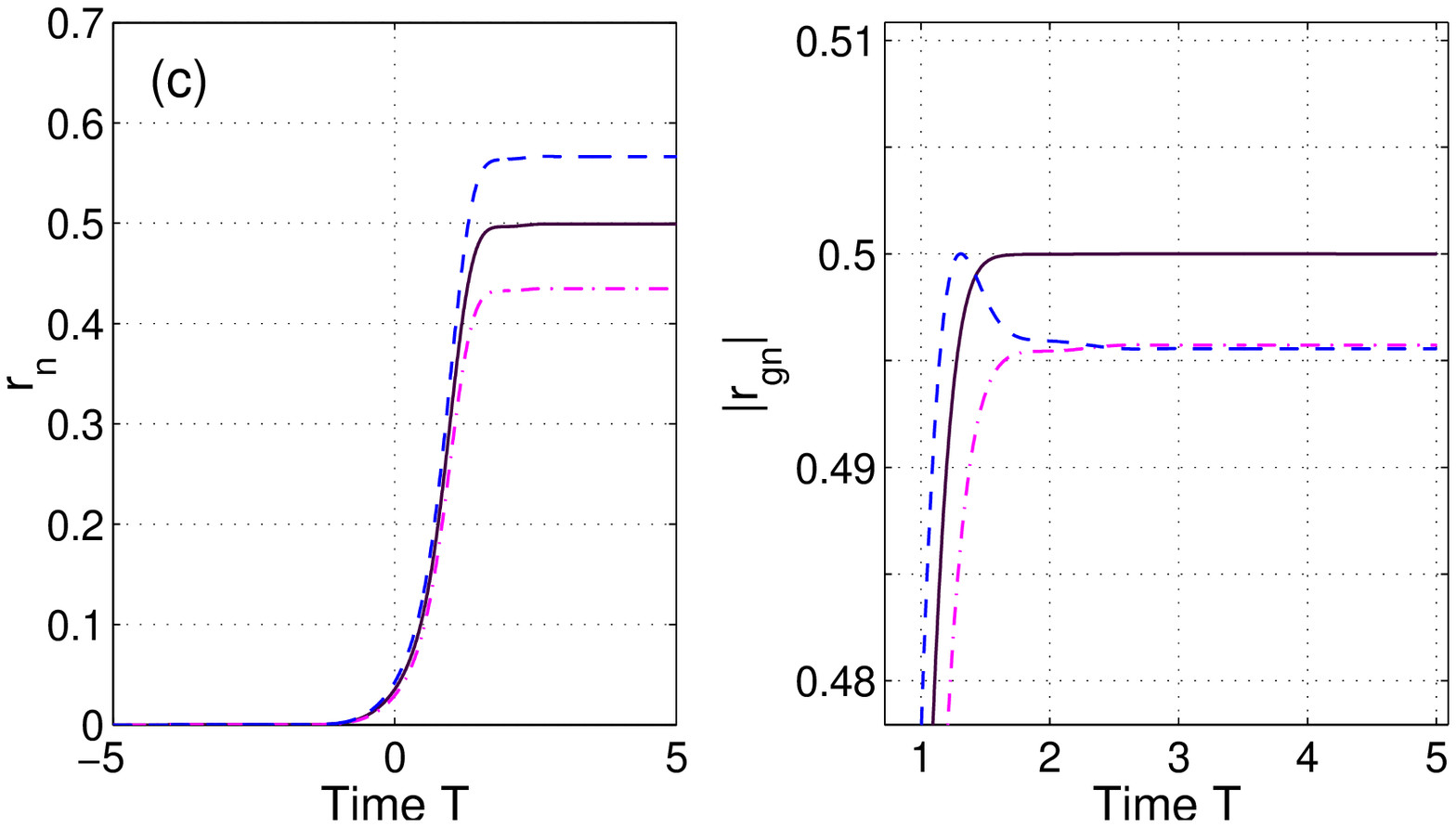}\\
\includegraphics[width=.5\columnwidth]{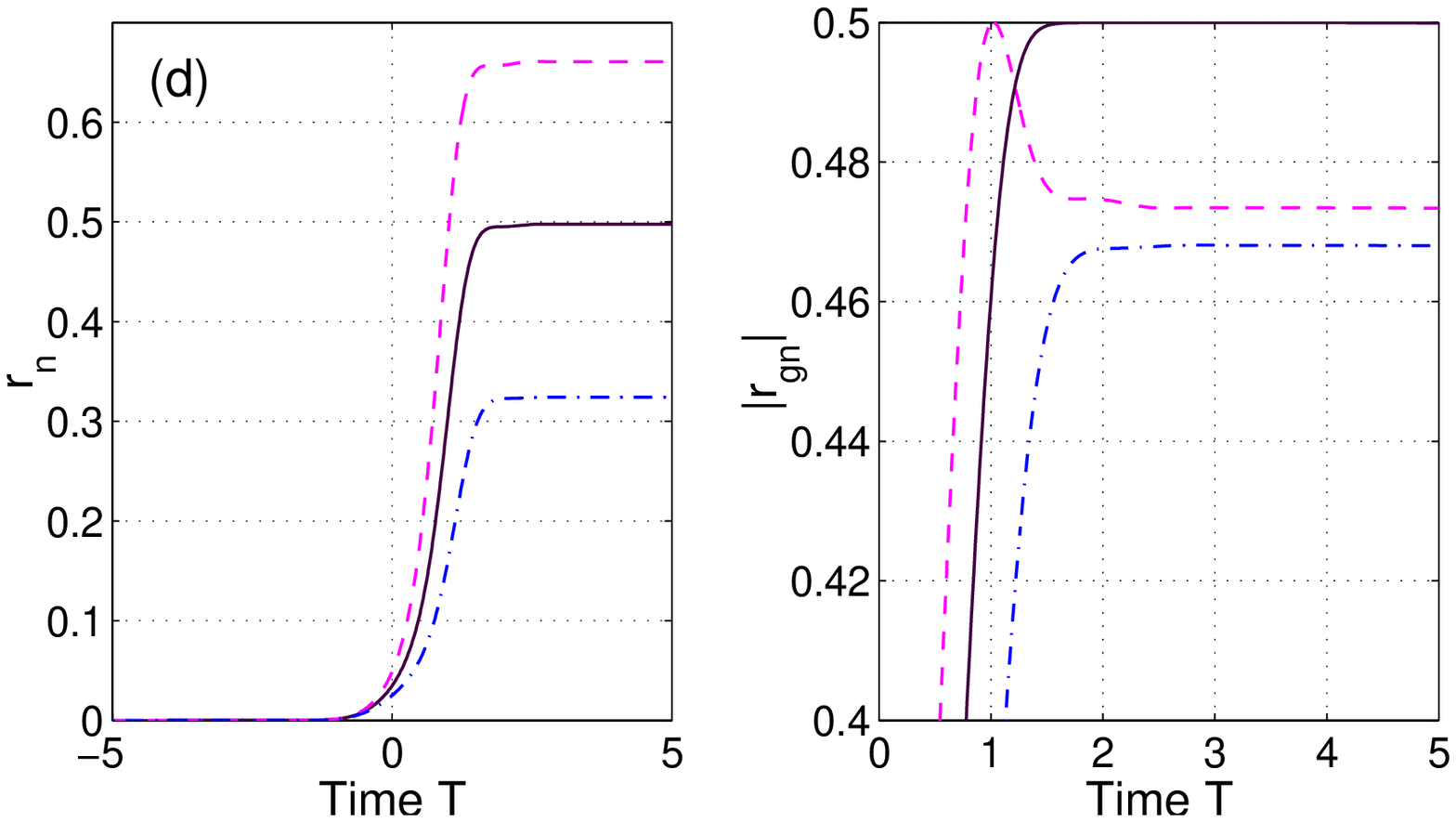}
\caption{\label{2lstrob} Robustness of maximum coherence created with the
Gaussian pulses of two-photon excitation and Stark shift of the resonance,
while the dynamic self-shift of the resonance is neglected ($\beta=0$), and
$\tau_{St}/\tau_1=1.6$. (a) -- $\delta=5$, $R=3.18$, $\delta\tau=1.8$, solid
-- $S=7.4$, dashed -- $S=6.7$, dash-dotted -- $S=8.1$; (b) -- $S=7.4$,
$R=3.18$, $\delta\tau=1.8$, solid line -- $\delta=5$, dashed --
$\delta=4.5$, dash-dotted -- $\delta=5.5$; (c) -- $S=23$, $\delta=20$,
$\delta\tau=1.34$, solid -- $R=3.48$, dashed -- $R=3.85$, dash-dotted --
$R=3.15$; (d) -- $R=3.48$, $S=23$, $\delta=20$, solid -- $\delta\tau=1.34$,
dashed -- $\delta\tau=1.2$, dash-dotted -- $\delta\tau=1.5$.}
\end{center}
\end{figure}
\begin{figure}
\begin{center}
\includegraphics[width=.5\columnwidth]{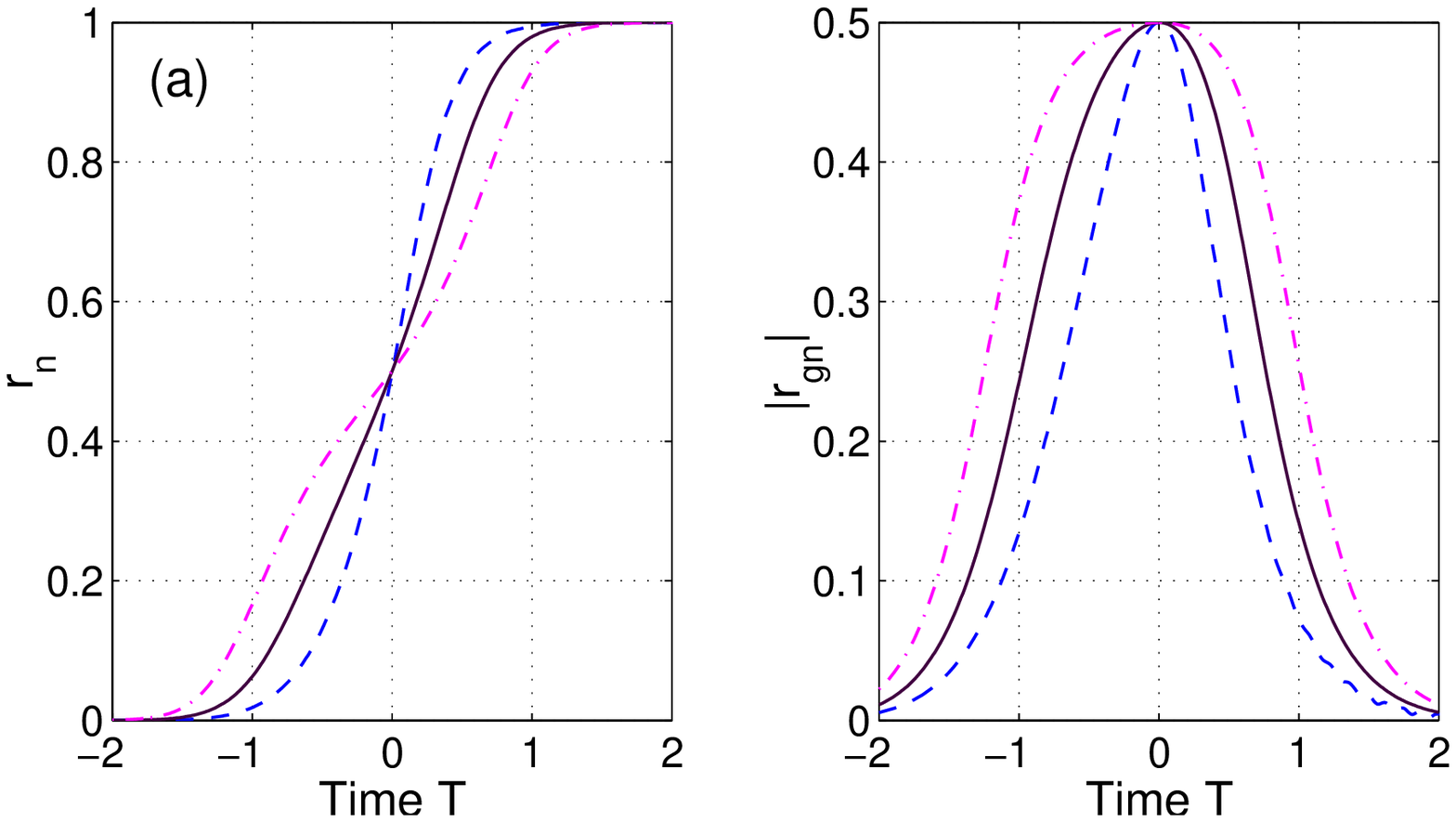}\\
\includegraphics[width=.5\columnwidth]{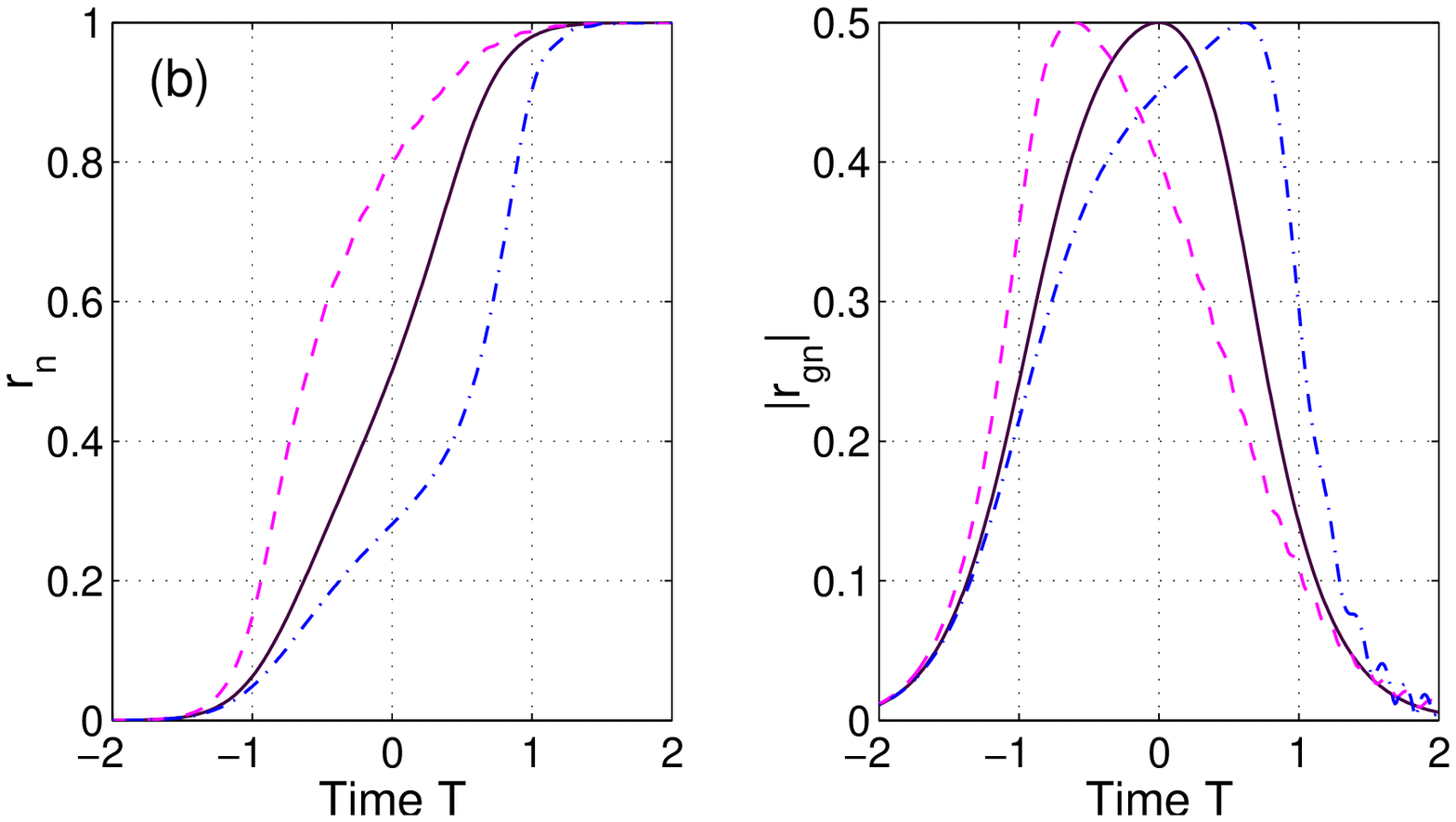}\\
\includegraphics[width=.5\columnwidth]{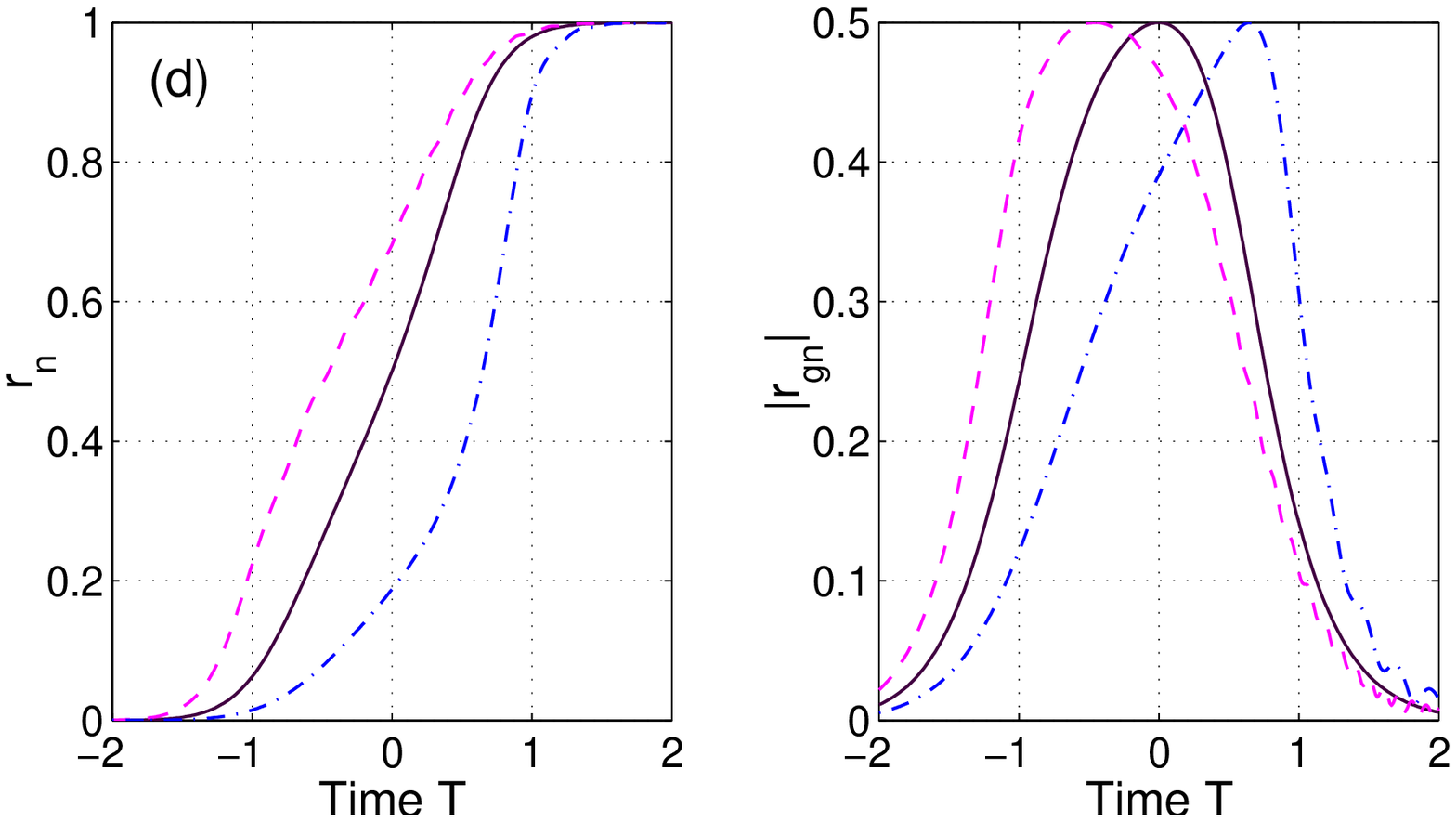}\\
\includegraphics[width=.5\columnwidth]{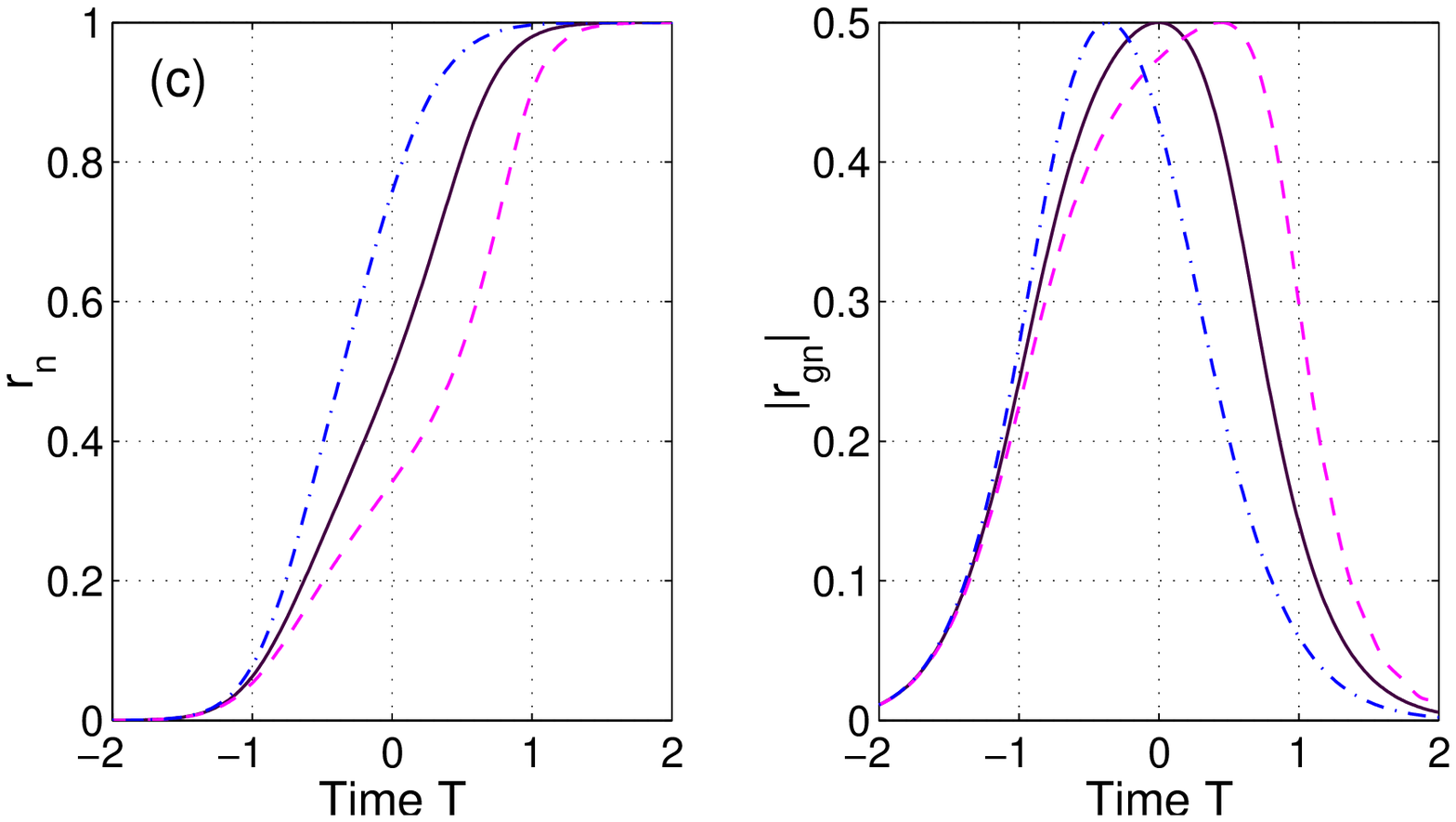}
\caption{\label{2lscrap} Rapid adiabatic passage assisted by Stark-chirp and
two-photon coherence, while the dynamic self-shift of the resonance is
ignored ($\beta=0$), and $\tau_{St}/\tau_1=1.6$. (a) -- $\delta=24$, $S=75$,
$\delta\tau=1.7$, solid line  -- $R=30$, dashed -- $R=15$, dash-dotted --
$R=60$; (b) -- $S=75$, $R=30$, $\delta=24$, solid -- $\delta\tau=1.7$,
dashed -- $\delta\tau=1.1$, dash-dotted -- $\delta\tau=2.3$; (c) --
$\delta=24$, $R=30$, $\delta\tau=1.7$, solid line -- $S=75$, dashed --
$S=50$, dash-dotted -- $S=150$; (d) -- $R=30$, $S=75$, $\delta\tau=1.7$,
solid line -- $\delta=24$, dashed -- $\delta=12$, dash-dotted --
$\delta=48$.}
\end{center}
\end{figure}
The results of simulations depicted in Fig. \ref{2lstrob} show that the
variation of the given radiative parameters about 10\% of their optimum
values leads to comparable change of the population transfer. As an
important outcome, it demonstrates substantially less change of the
coherence amplitude and therefore attests to its robustness. This is due to
the fact that the product of the probability amplitudes of the ground and
excited states changes less than the square modulus of probability amplitude
of the upper state.
\subsubsection{Stark-chirped rapid adiabatic passage and transient
maximum coherence}
As indicated above, the key idea of SCRAP \cite{Ric00} is to enhance the
transition probability at $t_1$ and to minimize it at $t_2$ (see
Eq.~(\ref{time})), which is possible if the evolution is adiabatic at $t_1$
and diabatic at $t_2$. Creation of maximum coherence is ensured in the
process of rapid adiabatic passage of the entire population from the ground
to excited state, where maximum coherence is always induced at instants when
populations of the level becomes equal. Figure \ref{2lscrap} presents such
behavior. It shows robustness of such passage, although the instants when
coherence reaches its maximum may change substantially.
\begin{figure}
\begin{center}
\includegraphics[width=.5\columnwidth]{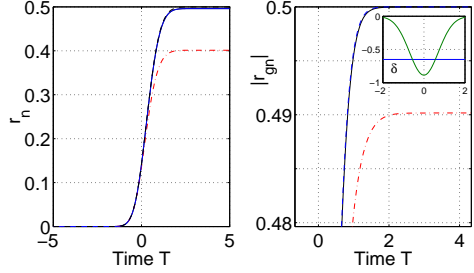}
\caption{\label{1selfst} Effect of dynamic self-shift of two-photon
resonance $s_1=\beta R$  on creation of persistent maximum coherence. $S=0$.
$R=0.886$, solid line -- $\beta=0$, $\delta=0$ ($\pi/2$-pulse), dash-dotted
-- $\beta=-1$, $\delta=0$, dashed (overlaps the solid one) -- $\beta=-1$,
$\delta=-0.65$. Insert -- dependence of the dynamic self-shift of the
two-photon resonance $s_1$ on time. The static deviation from the resonance
is marked by the straight line.}
\end{center}
\end{figure}
\begin{figure}
\begin{center}
\includegraphics[width=.5\columnwidth]{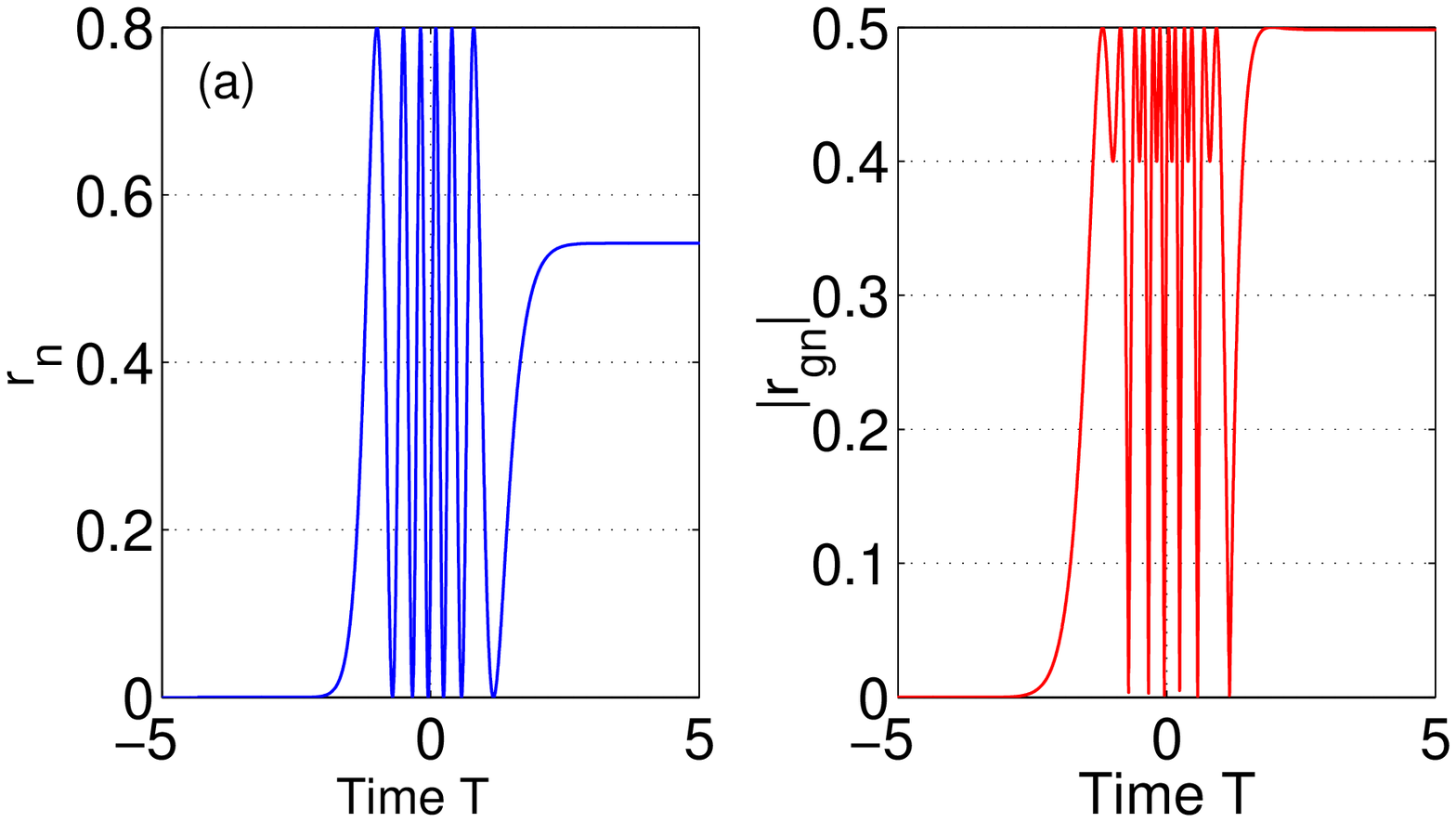}\\
\includegraphics[width=.5\columnwidth]{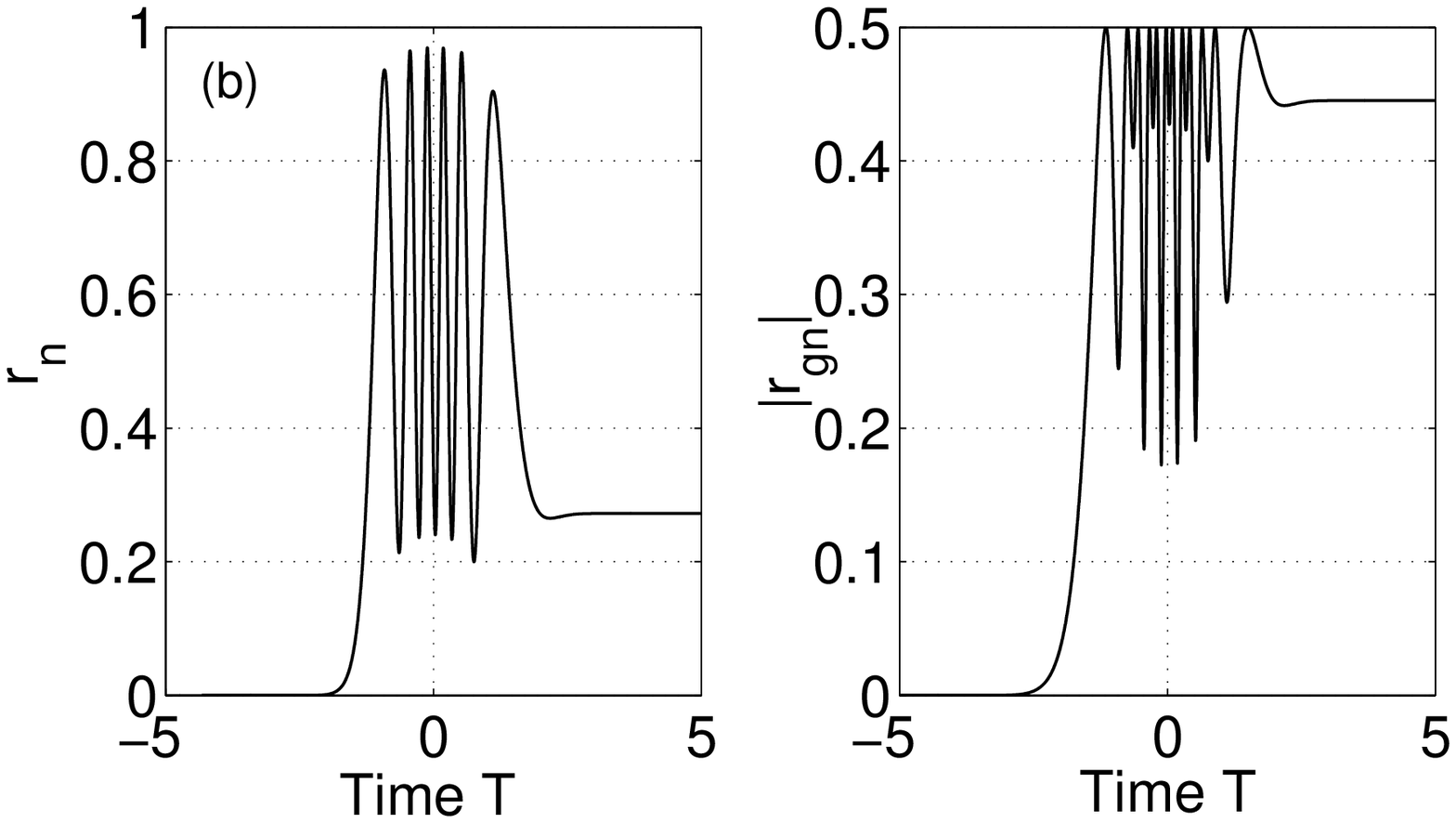}\\
\includegraphics[width=.5\columnwidth]{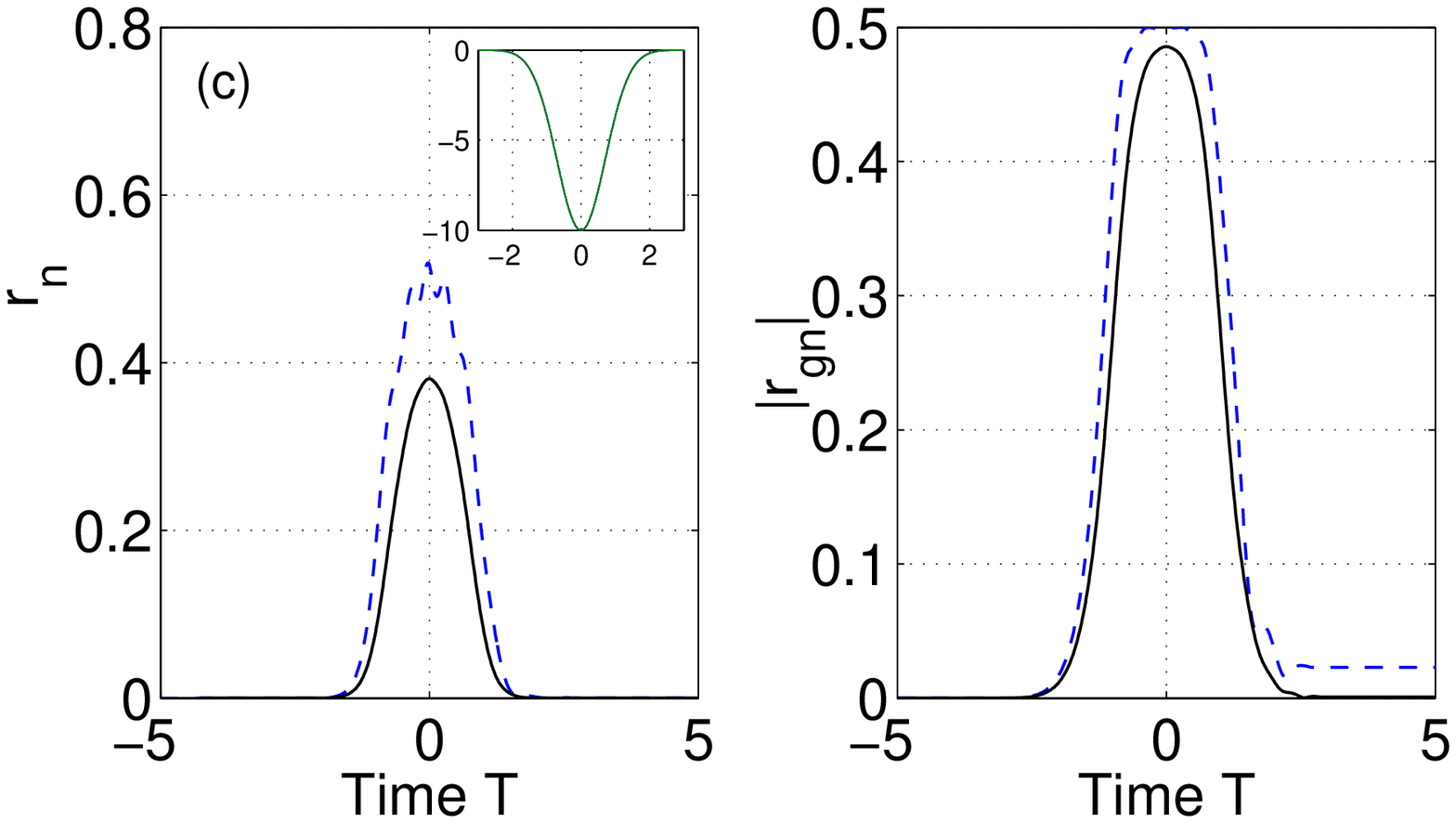}
\caption{\label{selfst1} Population $r_n$ and coherence $|r_{gn}|$
determined by the dynamic shift of the two-photon resonance driven both by
resonant excitation $s_1=\beta R$ and by an auxiliary Stark-shift field $s$.
$R=20$, $S=0$, $\beta=-0.5$; (a)-- $\delta=0$, (b)-- $\delta=-3$, (c) --
dashed line  -- $\delta=-10$, solid -- $\delta=-15$. Insert -- dependence of
the dynamic shift $s_1$ induced by two-photon excitation (dashed line) and
by two-photon excitation in cooperation with the auxiliary Stark field
$s_1+s$ (solid line).}
\end{center}
\end{figure}
\subsubsection{Maximum coherence and rapid adiabatic passage controlled by
the  dynamic shift of a two-photon resonance produced by both fundamental
and Stark Gaussian pulses}
In a multi-level system with appreciable difference between the principal
quantum numbers of the ground and excited states, two-photon excitation may
substantially contribute to the dynamic shift of a resonance. This makes
transient effects  and consequently the choice of the parameters ensuring
maximum coherence and rapid adiabatic passage more complicated. Figure
\ref{1selfst} presents such behavior, where the solid line shows feasibility
to  generate persistent maximum coherence in the absence of dynamic shift of
the resonance (a $\pi/2$ pulse). It is seen that even a relatively small
{\it self-induced} dynamic shift of the two-photon resonance may
substantially change the time-behavior of the system (dash-dot plot).
However, persistent maximum coherence can be restored through an
appropriately chosen static deviation from the resonance provided that the
Rabi frequency is around the value corresponding to a $\pi/2$-pulse.

Under larger intensities of two-photon excitation, the dynamics of the
system controlled by the dynamic self-shift becomes more complicated, so
that generation of persistent coherence without the means of independent
control of the shift with an auxiliary pulse may become impossible (Fig.
\ref{selfst1}).
\begin{figure}[!h]
\begin{center}
\includegraphics[width=.5\columnwidth]{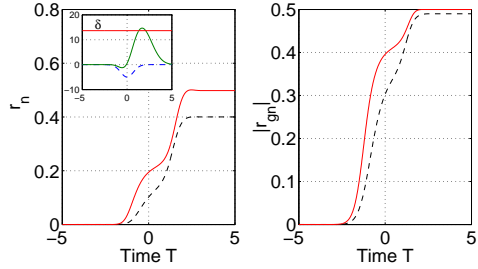}
\caption{\label{perscoh2st} Creation of persistent maximum coherence
controlled by appropriate static deviation from resonance and by appropriate
delay of he auxiliary Stark-shift pulse in the system experiencing the
dynamic self-shift of the two-photon resonance. $s_1=\beta R$, $\beta=-0.5$,
$S=15$, $\delta=13.65$, $\delta\tau=1.6$; dashed line -- $R=15$, solid --
$R=30$. Insert -- dependence of the dynamic shift $s_1$ induced by
two-photon excitation (dashed line) and by two-photon excitation in
cooperation with the auxiliary Stark field $s_1+s$ (solid line). The static
deviation from the resonance is marked by the straight line.}
\end{center}
\end{figure}
Indeed, independent control of the dynamic shift of the two-photon resonance
with the aid of the auxiliary Stark-shift pulse,  and based on the above
discussed algorithm,  ensures creation of persistent maximum two-photon
coherence (Fig. \ref{perscoh2st}).
\begin{figure}[!h]
\begin{center}
\includegraphics[width=.5\columnwidth]{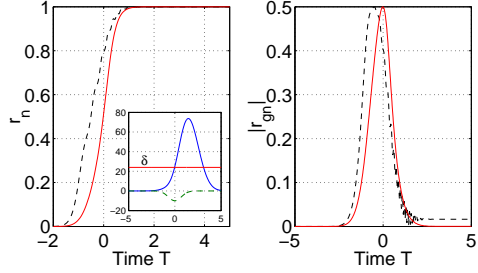}
\caption{\label{selfscr} Persistent Stark-chirp-assisted rapid adiabatic
passage controlled by the appropriately delayed pulses and by static
deviation from resonance in the system experiencing a dynamic self-shift of
the two-photon resonance. $s_1=\beta R$, $\beta=-0.5$, $S=75$, $R=20$,
$\delta\tau=1.4$; dashed line -- $\delta=10$, solid -- $\delta=24$. Insert
-- dependence of the dynamic shift $s_1$ induced by two-photon excitation
(dashed line) and by two-photon excitation in cooperation with the auxiliary
Stark field $s_1+s$ (solid). the static deviation from the resonance is
marked by the straight line. }
\end{center}
\end{figure}

Figure \ref{selfscr} illustrates similar opportunities for {\it transient}
maximum coherence generated through persistent transfer of the entire
population of ground state to the upper state (SCRAP).
\section{Dynamics of energy level populations, four-wave mixing
at maximum coherence, and  VUV generation in Hg vapors}
In this section, we apply our theory for simulating the features of
two-photon resonant FWM controlled by SCRAP with the aid of a numerical
model appropriate for quantum transitions in Hg. Then the energy levels
depicted in (Fig.~\ref{4lsh}) are attributed to those of $Hg$ as follows:
$g$ -- $6s^2\,^1S_0$, $n$ -- $8s\,^1S_0$,  $m$ -- $6p\,^1P_1$, $l$ --
$6p\,^1P_1$, $f$ -- $7p\,^1P_1$, and the laser parameters are similar to
those described in Ref. \cite{Ric00}: $\lambda_1=268.8$~nm,
$\lambda_{St}=1064$~nm, $\lambda=532$~nm, $\tau_1=3$~ns. Then
$\lambda_{-}=179.8$~nm, $\lambda_{+}=107.3$~nm,
$\delta_{nf}\sim\delta=8.9\times10^3$, $\delta_{nl}= -2.2 \times10^4$ and
$\delta_{gm}=-2.4\times10^5$ (all detunings are scaled to $\tau_1^{-1}$).
The parameters $a$ and $K_j$ are taken as $a=0.345$, $K_-=1, K_2=0.04,
K_1=0.67$. We have assumed that other levels, which are not depicted here,
do not contribute substantially to the coupling. The other pulse durations
are chosen as: $\tau_2=\tau_1$, $\tau_{St}/\tau_1=1.6$. The delay between
all these pulses can be varied. We introduce the parameters
$G_i=|g_i|\tau_1/2\pi$ and $Z=\xi\tau_1/2\pi$, which denote  reduced by
$2\pi$ the number of Rabi oscillations during the pulse $E_1$ and the medium
length scaled to the resonant absorption length at the transition $gl$ for
the nonmonochromatic radiation with the spectral width $\tau_1^{-1}$. The
results of the simulations are presented below.

\begin{figure}[!h]
\begin{center}
\includegraphics[width=.5\columnwidth]{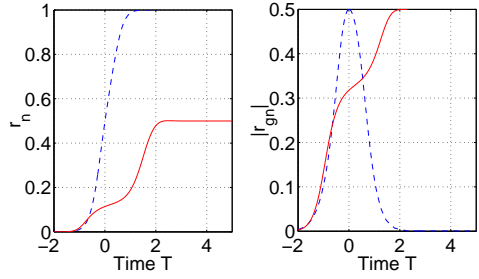}
 \caption{\label{rbegin} Dependence of the population $r_n$ and
coherence $|r_{gn}|$ on time at the entrance to the medium. $G_{01}=910$,
$G_{0st}=325$. Solid line -- $\delta=0.5$, $\delta\tau_{st}=-3$, dashed --
$\delta=5.6$, $\delta\tau_{st}=2$. Other parameters are given above.}
\end{center}
\end{figure}

\begin{figure}[!h]
\begin{center}
\includegraphics[width=.5\columnwidth]{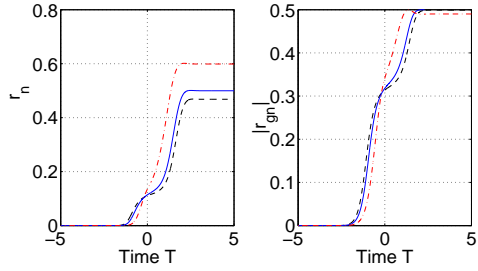}
 \caption{\label{entr1} Dependence of the population $r_n$ and
coherence $|r_{gn}|$ on time at the entrance to the medium. $G_{0st}=325$,
$\delta=0.5$, $\delta\tau_{st}=-3$. Solid line -- $G_{01}=910$, dashed --
$G_{01}=1173$, dash-dotted -- $G_{01}=525$.}
\end{center}
\end{figure}

\begin{figure}[!h]
\begin{center}
\includegraphics[width=.5\columnwidth]{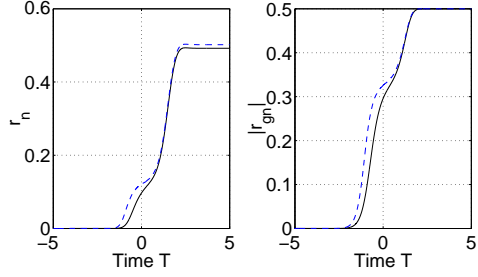}
\caption{\label{entr2} Dependence of the population $r_n$ and coherence
$|r_{gn}|$ on time at the entrance to the medium. $G_{01}=910$,
$\delta=0.5$, $\delta\tau_{st}=-3$. Solid line -- $G_{0st}=461$, dashed --
$G_{0st}=266$.}
\end{center}
\end{figure}

\begin{figure}[!h]
\begin{center}
\includegraphics[width=.5\columnwidth]{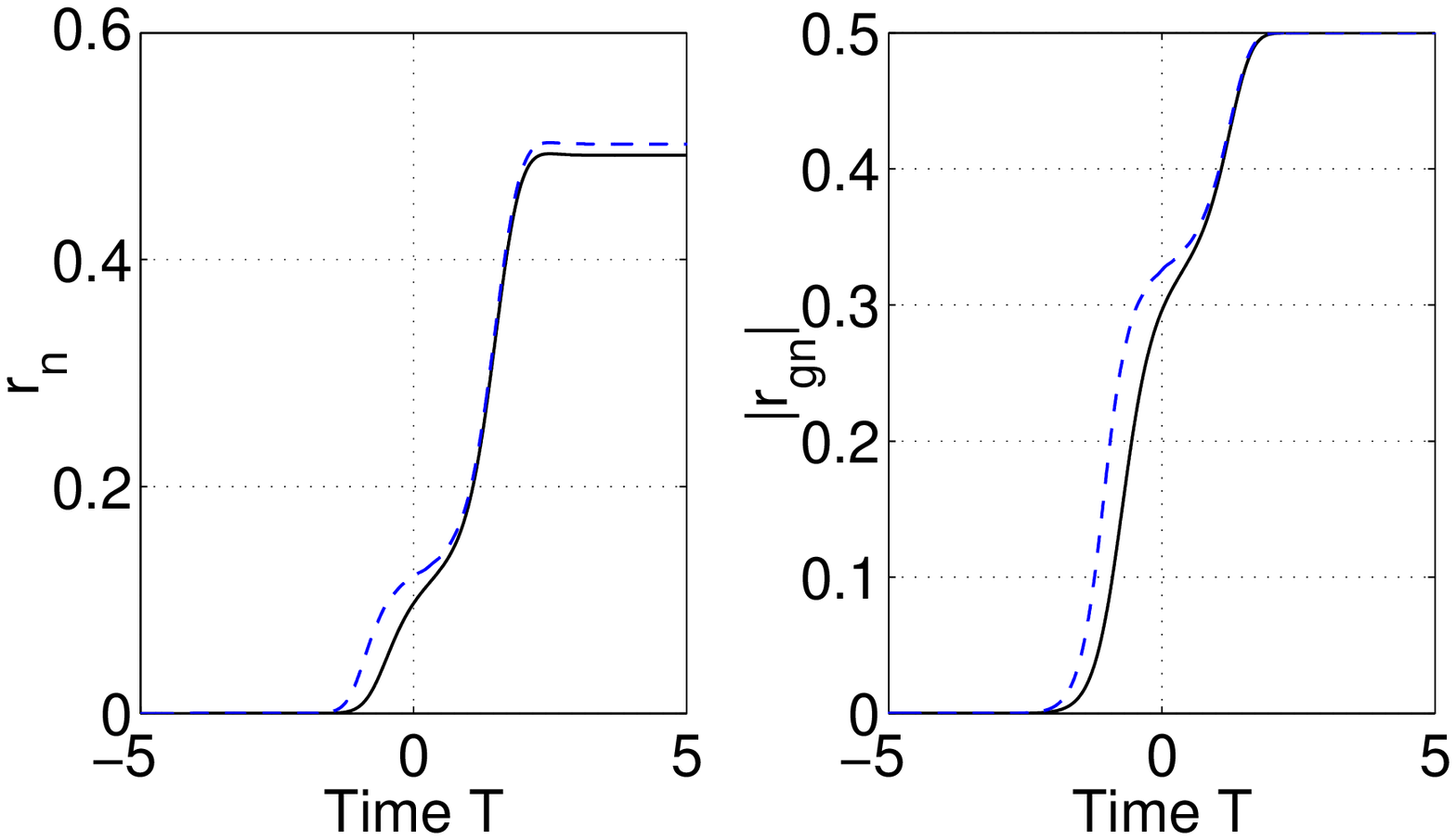}
 \caption{\label{entr3} Dependence of the population $r_n$ and
coherence $|r_{gn}|$ on time at the entrance to the medium. $G_{01}=910$,
$G_{0st}=325$, $\delta=0.5$. Solid line -- $\delta\tau_{st}=-4$, dashed --
$\delta\tau_{st}=-3$, dash-dotted -- $\delta\tau_{st}=-2$. }
\end{center}
\end{figure}

As outlined above, in order to achieve  a transfer efficiency close to
unity, one must fulfill the adiabatic condition described in Ref.
\cite{Ric00}). Through a proper choice of $G_{01}$, $G_{0st}$, static
detuning $\delta$ and pulse delay $\delta\tau_{st}$, we can ensure various
values and evolution in time of the populations and coherence at the
entrance to the medium. Figures \ref{rbegin}-\ref{entr3} demonstrate the
possible dynamics leading either to maximum population transfer  or to equal
populations and maximum coherence. These pictures (along with
Fig.~\ref{2lscrap}) illustrate the robustness of these processes, regardless
of the significant change in some of the parameters.

\begin{figure}[h]
\begin{center}
\includegraphics[width=.4\columnwidth]{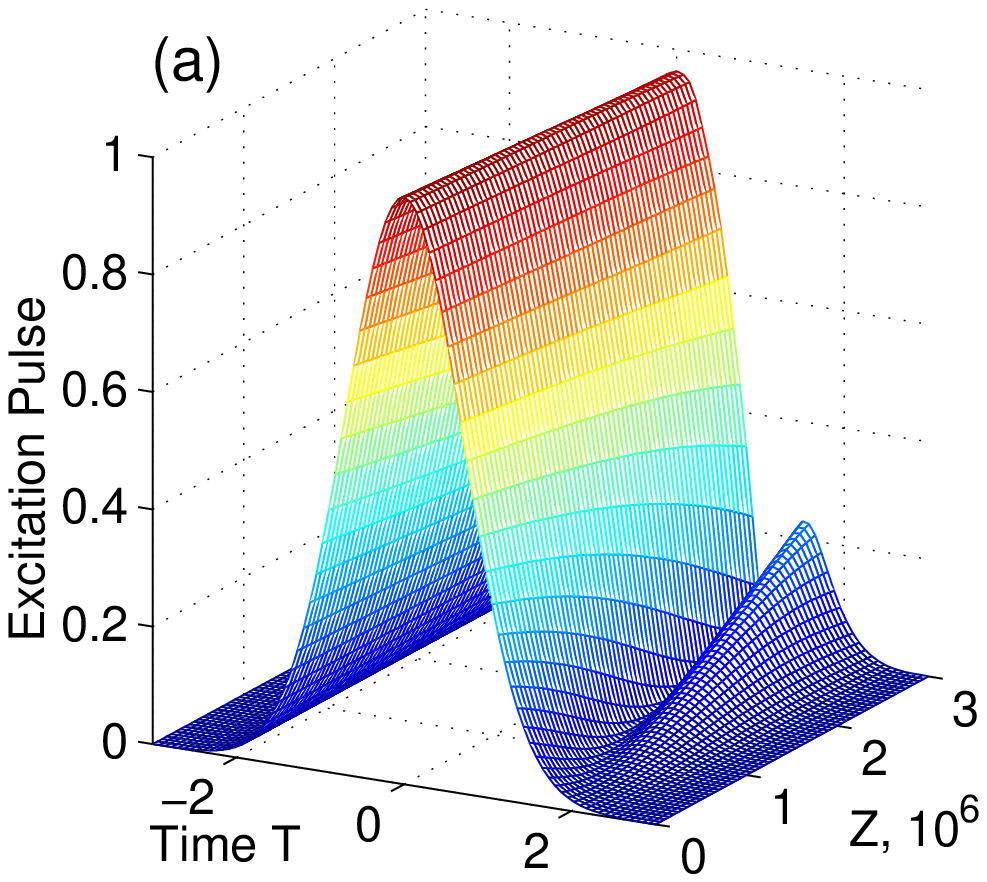}
\includegraphics[width=.4\columnwidth]{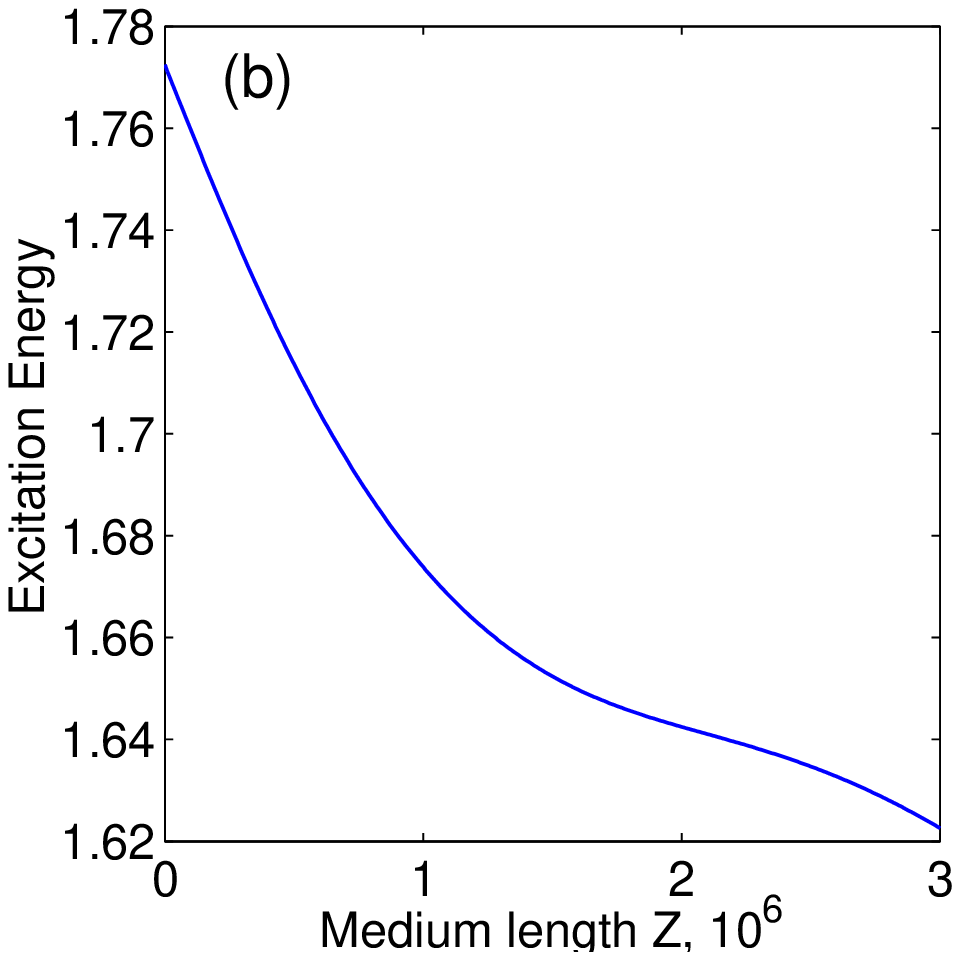}
 \caption{\label{3DMshapegs} Evolution of the shape of the pulse
$|G_1/G_{01}|^2$ (a) and its energy $\int |G_1/G_{01}|^2 dt$ (b)
along the medium. $\delta=0.5$, $\delta\tau_{st}=-3$.}
\end{center}
\end{figure}

The generated radiation depends strongly on the dynamics of the coherence
and on the delay at which the pulse $G_2$ to be converted is applied. For
the cases depicted in Fig.~\ref{rbegin}, one can choose either a relatively
short interval, when populations of the levels $n$ and $g$ becomes equal and
the coherence amplitude reaches its maximum, or a 'plateau' with lower but
almost constant amplitude of the two-photon coherence $|r_{gn}|$ in time.
Furthermore, in our simulations, if it is not indicated otherwise, we assume
that $G_{0st}=325$ and at the entrance of the Hg cell, $G_{01}=910$ (which
corresponds to $R=15$, $S=75$), and the amplitude of the pulse $G_2$ is
chosen to be non-perturbatively small ($G_{20}=1.6\times 10^{-2}$). The
delay of the pulse $G_2$ is set as $\delta\tau_2=5$, so that this pulse
overlaps the plateau in the time dependence of the coherence $|r_{gn}|$ at
the entrance to the medium (Fig.~\ref{rbegin}, solid line), or
$\delta\tau_2=0$, so that the maximum of the convertible pulse coincides
with the maximum of the time dependence of the coherence (Fig.~\ref{rbegin},
dashed line).
\begin{figure}
\begin{center}
\includegraphics[width=.25\columnwidth]{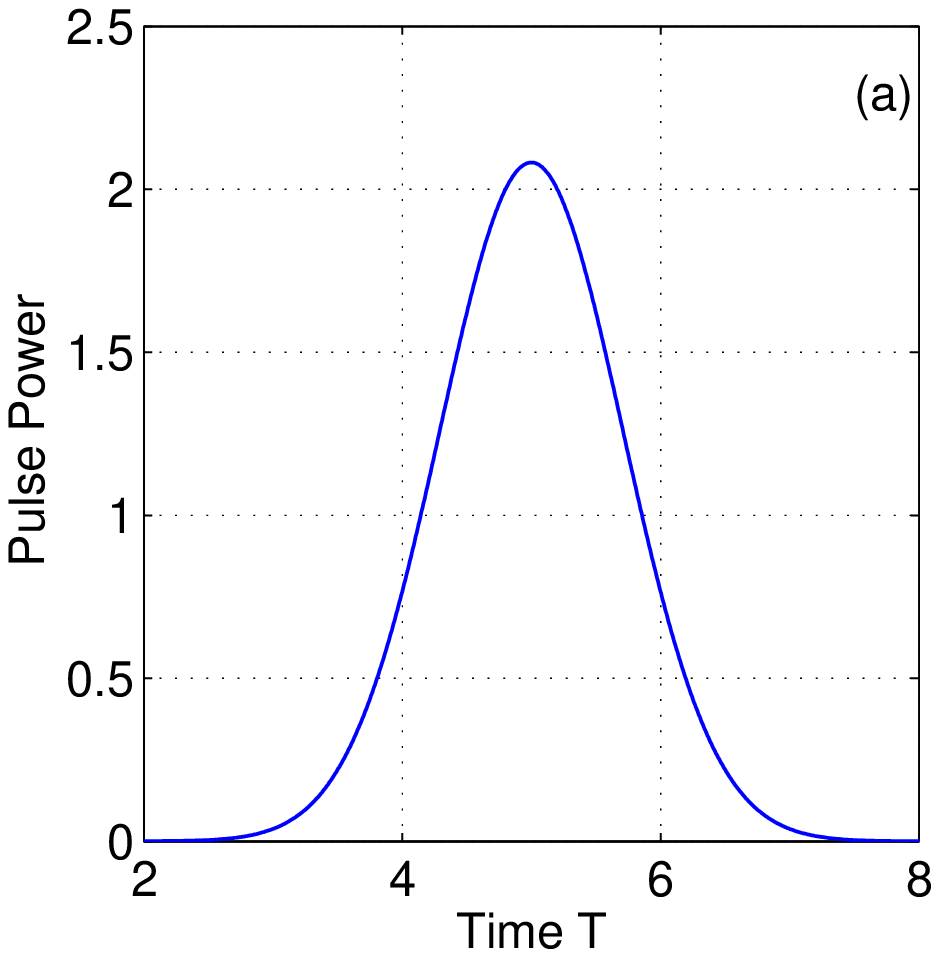}
\includegraphics[width=.25\columnwidth]{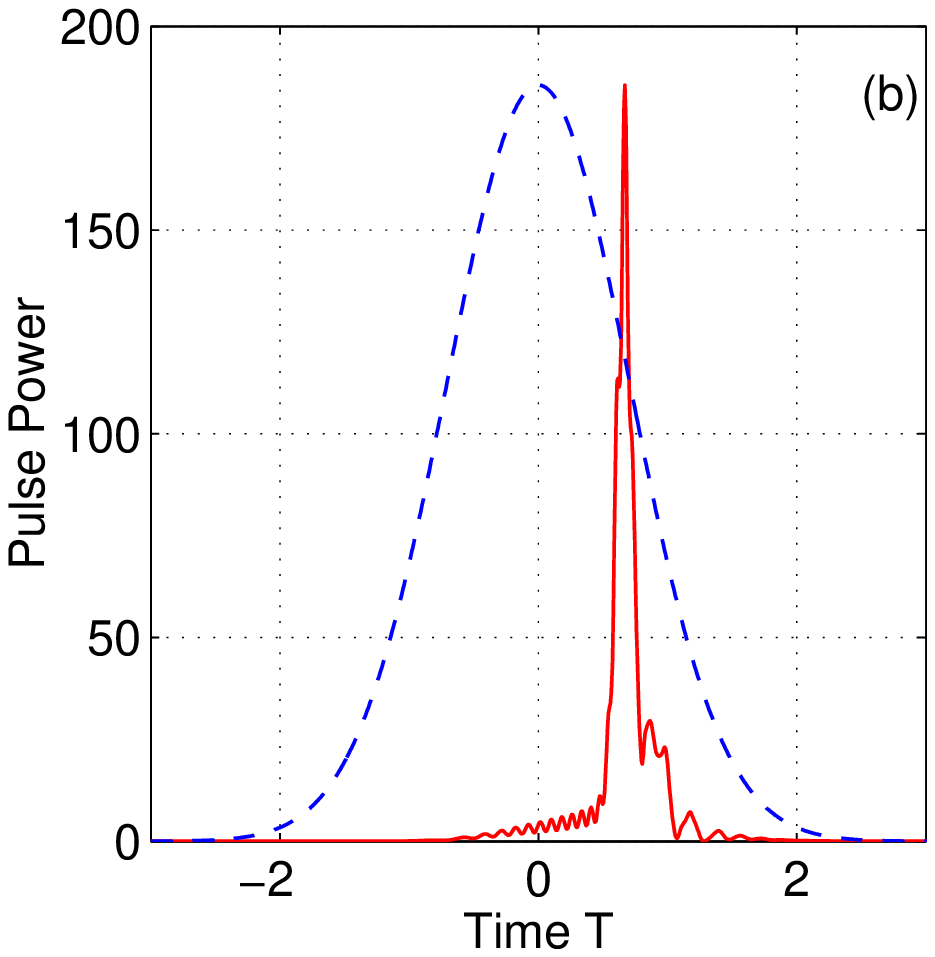}\\
\includegraphics[width=.25\columnwidth]{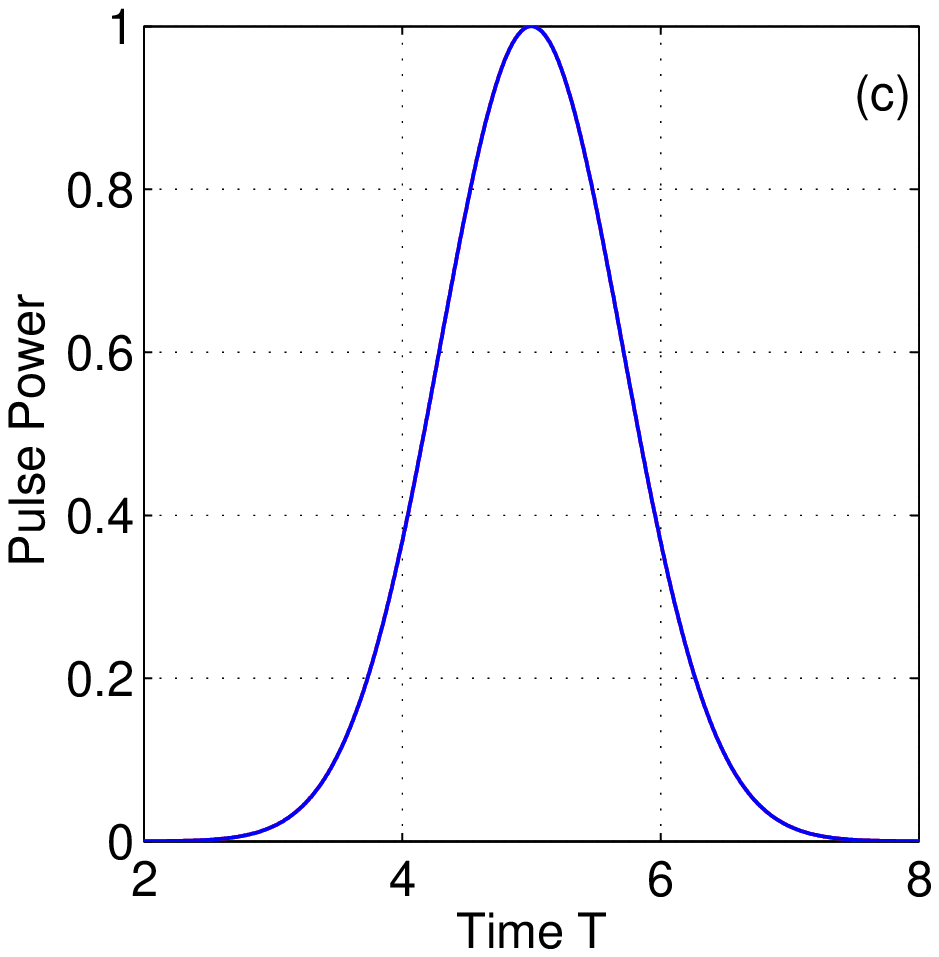}
\includegraphics[width=.24\columnwidth]{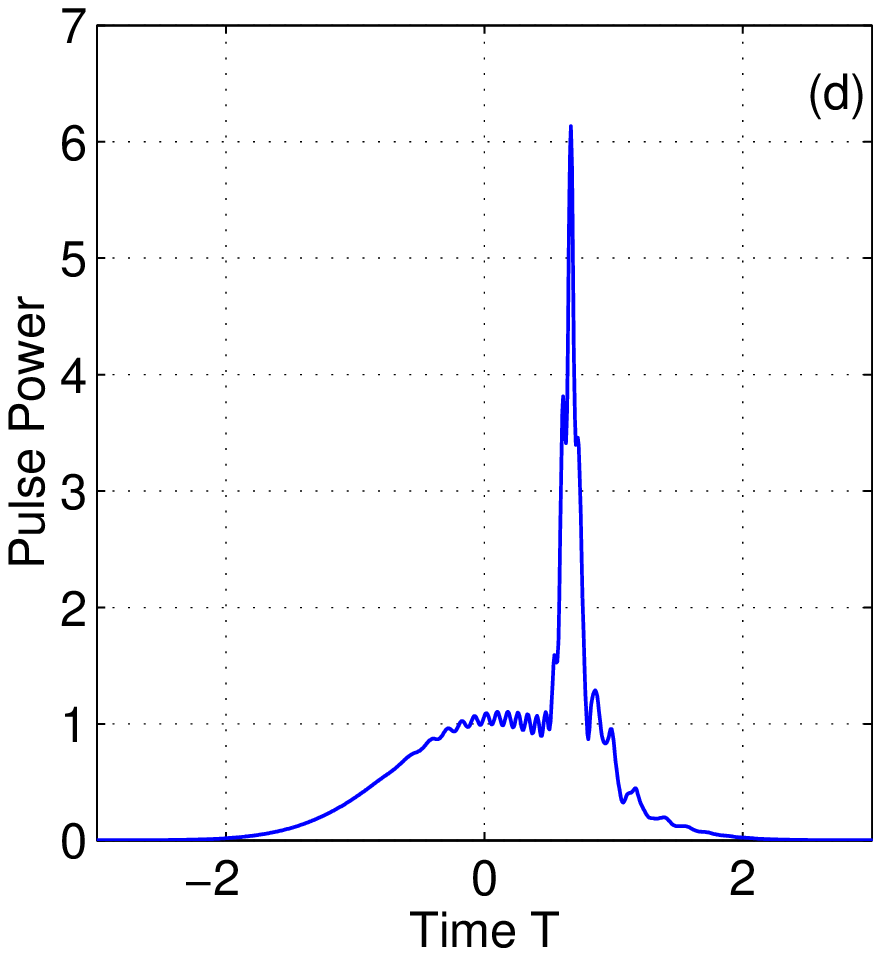}
\caption{\label{shapegs} Shapes of the generated $|G_-/G_{02}|^2$ (a,b) and
amplified $|G_2/G_{02}|^2$ (c,d) pulses at $Z=3\times 10^6$. (a) and (c)
correspond to the plots in Fig. \ref{rbegin} (solid line), with
$\delta\tau_2=5$, and (b) and (d) to the plots at Fig. \ref{rbegin} (dashed
line) with $\delta\tau_2=0$. The dashed line in plot (b) is the input pulse
scaled to the maximum of the output one. For the plot (c), the shapes of the
input and output pulses are similar.}
\end{center}
\end{figure}
Figures \ref{shapegs}(a-d) computed for difference-frequency processes show
that the shape of the generated and convertible pulses may substantially
vary along the medium, subject to the coherence dynamics and the time delay
$\delta\tau$ at the entrance of the medium. It is seen that if the pulse
$G_2$ overlaps the plateau of the time dependence of $|r_{gn}|$, there is no
significant transformation of the pulse shape along the medium (the
corresponding plots in Fig.~\ref{shapegs}(c) coincide). On the contrary, if
the maximum coherence is reached in a transient regime, and the maximum of
the pulse $G_2$ coincides with the instant when the populations of the
levels $r_g$ and $r_n$ become equal at the entrance to the medium, the
change of pulse shapes is most significant (Fig.~\ref{shapegs}(b,d)). In the
first case, all the parts of the pulse $G_2$ are converted homogeneously. In
the second case, the wings of the pulse $G_2$ are not converted and enhanced
since the shape of the pulse $|r_{gn}|$ is substantially narrower than that
of $G_2$. This leads to a sharpening of both the generated and convertible
(i.e., amplified) pulses. The shape of the excitation pulse may
substantially change along the medium too, although its energy has
practically no change (see Fig.~\ref{3DMshapegs}).

\begin{figure}[!h]
\begin{center}
\includegraphics[width=.25\columnwidth]{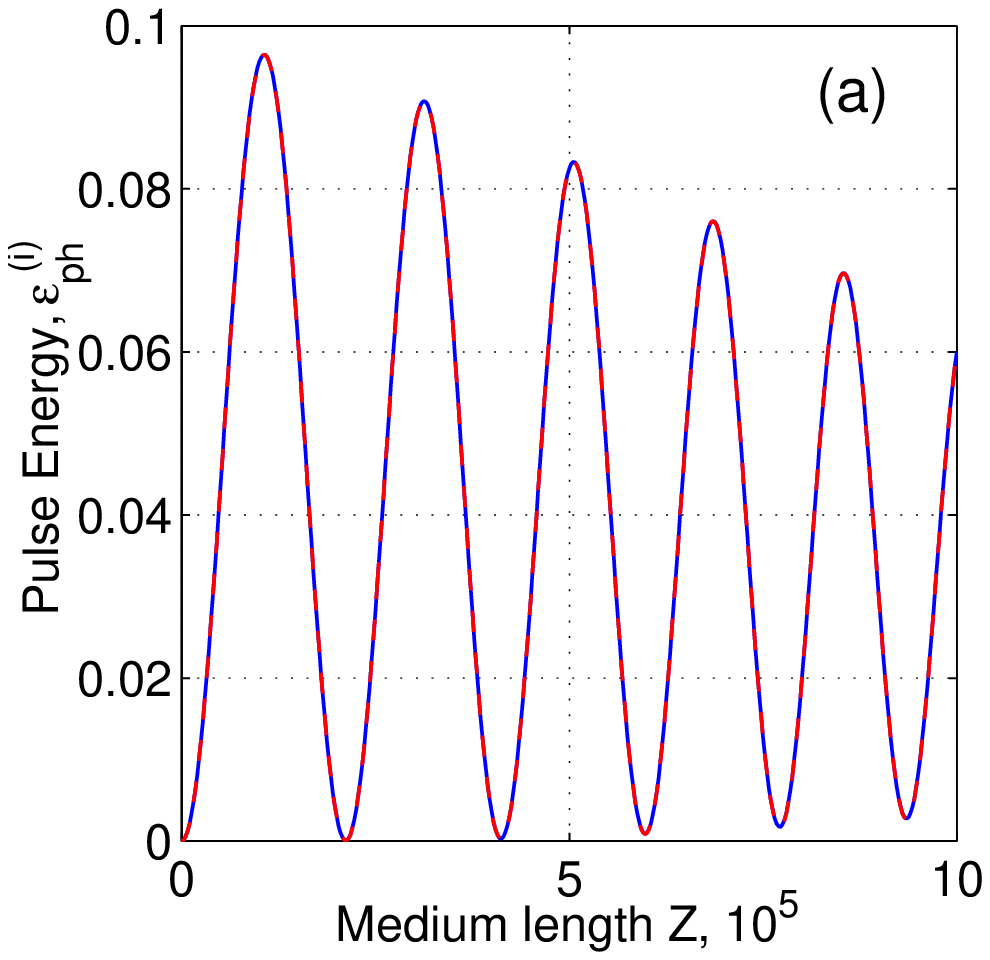}
\includegraphics[width=.25\columnwidth]{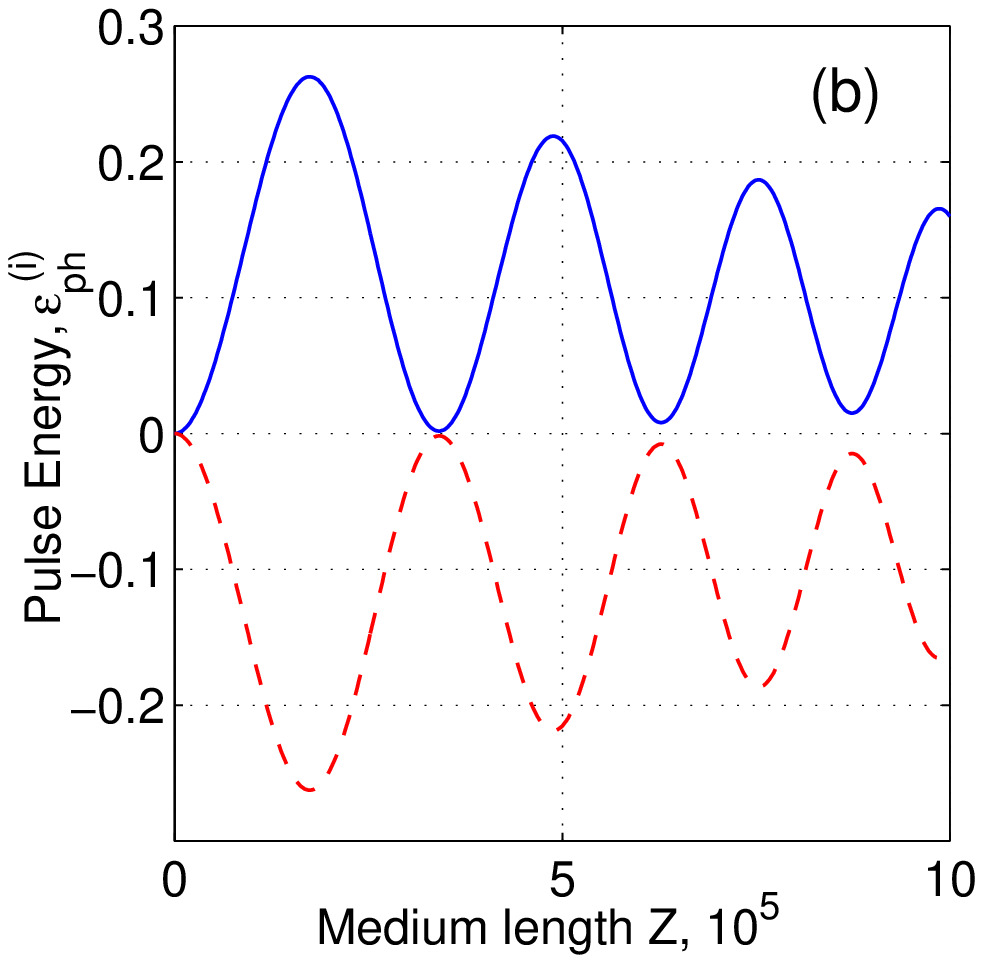}\\
\includegraphics[width=.25\columnwidth]{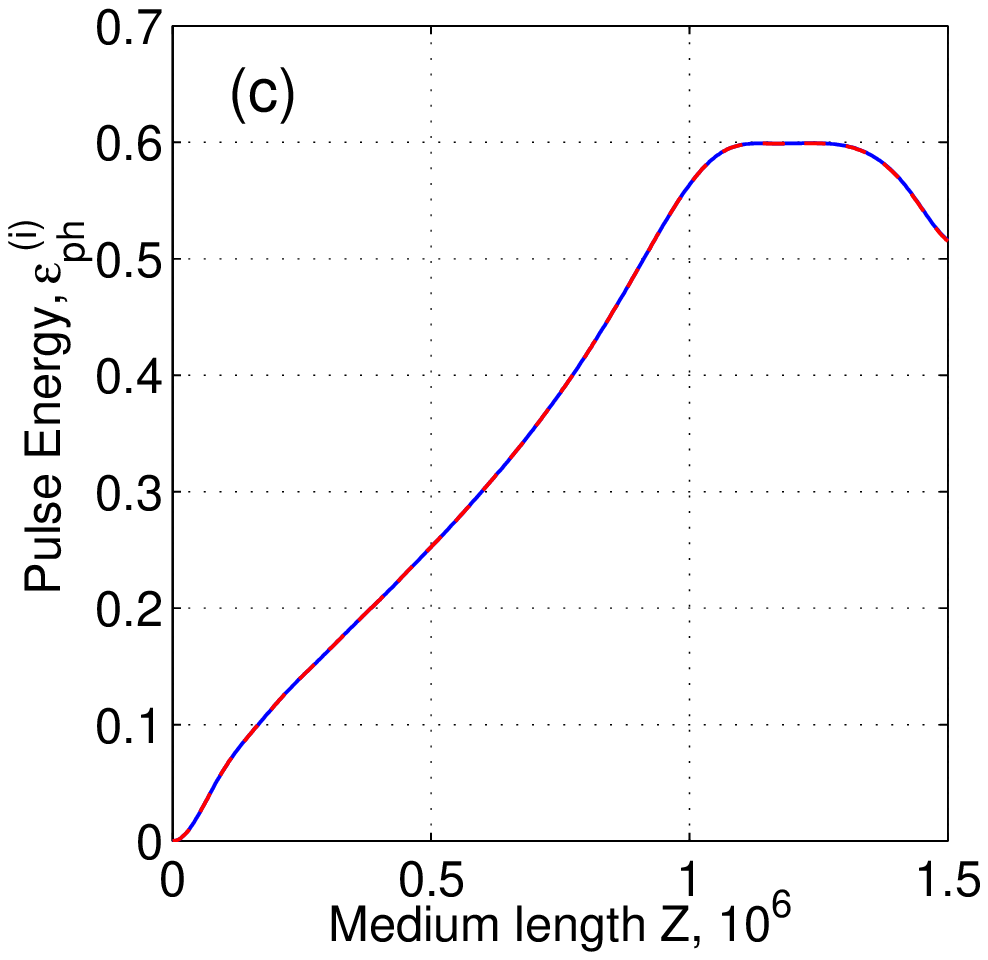}
\includegraphics[width=.25\columnwidth]{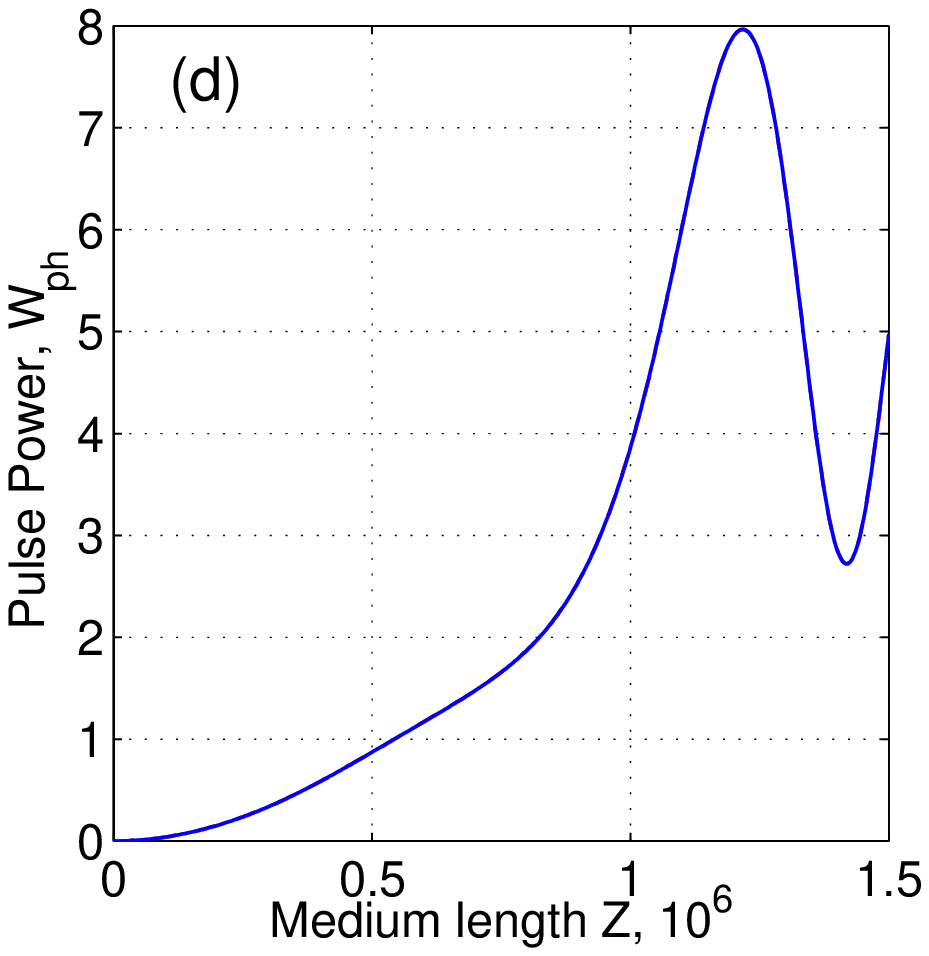}
\caption{\label{gdownz} Evolution  the pulse energy $\varepsilon_{ph}$ of
the convertible  and generated  fields, scaled to the initial number of
photons at convertible frequency $\omega_2$ along the medium. (a-c). (a) --
down-conversion $\omega_{-}=2\omega_1-\omega_2$ and (b) -- up-conversion
$\omega_{+}=2\omega_1+\omega_2$, both correspond to the plots at Fig.
\ref{rbegin} (solid line) whith  $\delta\tau_2=5$, and (c,d) to down
conversion, which correspond to the plots at Fig. \ref{rbegin} (dashed line)
whith $\delta\tau_2=0$. (c) -- energy $\epsilon_{ph}$, (d) -- peak pulse
power $W_{ph}^{max}$. For up-conversion (b) plots for generated (solid line)
and for convertible (dashed) fields are different, for down-conversion
(a,c,d) their behavior is similar.}
\end{center}
\end{figure}
\begin{figure}[!h]
\begin{center}
\includegraphics[width=.5\columnwidth]{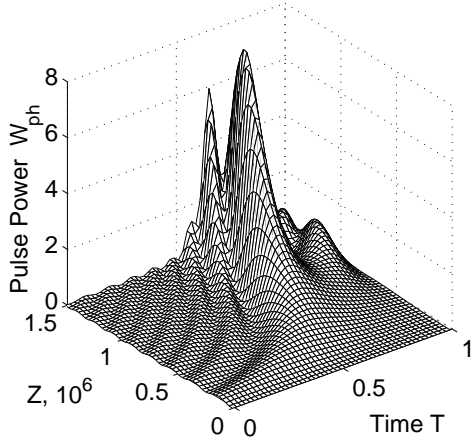}
\caption{\label{shpg2} Evolution  of the pulse shape $W_{ph}(T)$ of the
generated and convertible radiations (coincide), with all parameters the
same as in Fig. \ref{gdownz}(c).}
\end{center}
\end{figure}
\begin{figure}[!h]
\begin{center}
\includegraphics[width=.5\columnwidth]{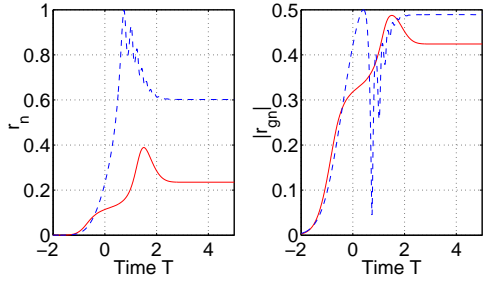}
\caption{\label{rins} Evolution in time of the population $r_n$ and
coherence $|r_{gn}|$ at the medium  length $Z=1\times10^6 $. $G_{01}=910$,
$G_{0st}=325$. Solid line -- $\delta=0.5$, $\delta\tau_{st}=-3$, dashed --
$\delta=5.6$, $\delta\tau_{st}=2$.}
\end{center}
\end{figure}
Figure \ref{gdownz}(a-c) shows the dependencies of the pulse energy
\begin{equation}\label{e}
\varepsilon_{ph}^{(i)}=\frac{\int|G_i(\tau)|^2\,d\tau-
\int|G_{0i}(\tau)|^2\,d\tau}{\int|G_{02}(\tau)|^2\,d\tau}
\cdot\frac{\omega_2|d_{ln}|^2}{\omega_i |d_{i}|^2},
\end{equation}
and Figure \ref{gdownz}(d) the pulse power at the instances when it reaches
its maximum
\begin{equation}\label{p}
W_{ph}^{(i)}=\frac{|G_i(\tau)|_{max}^2-|G_{0i}
(\tau)|_{max}^2}{|G_{02}(\tau)|_{max}^2}
\cdot\frac{\omega_2|d_{ln}|^2}{\omega_i |d_{i}|^2},
\end{equation}
along the medium. Both are scaled to the corresponding value for the
convertible field $E_2$ at the entrance of the medium. The plots (a,b) are
computed for the case when, at the entrance to the media, the pulse $E_2$
overlaps the plateau in the time dependence of the coherence $|r_{gn}|$. The
simulations show that in this case there is no difference between the plots
for energy and power. On the contrary, in the case when the second pulse
overlaps maximum of the transient coherence, the difference between energy
and power becomes significant (Fig.~\ref{gdownz}(c,d)), because pulse shapes
experience change along the medium (Fig.~\ref{shpg2}).

The figures show that the evolution of the fields along the medium is
strongly determined by the coherent nonlinear coupling. The difference in
evolution of the field $E_2$ in plots (a,b) is due to the fact that the
field $E_2$ is enhanced in the difference-frequency mixing process and, by
contrast, depletes during the course of the sum-mixing process. The
simulation reveals a substantial difference in the evolution, depending on
such coupling parameters at the entrance to the medium, like the static
detuning and delays between pulses, which control the coherence in time and
along the medium. The maximum achieved amplitude of the generated field may
change by several times, and the conversion efficiency for the sum-frequency
process may approach unity.
The figures prove the feasibilities of substantial (by more than several
tens times) improvement of the conversion efficiency  by a judicious fit of
the parameters at the medium entrance. Since the pulse shape of the driving
field $E_1$ varies along the medium, consequently, the populations $r_g$,
$r_n$, as well as the coherence $r_{gn}$, vary along the medium too.
Therefore, the system being optimally prepared at the entrance
(Fig.~\ref{rbegin}) may evolve to less optimum along the medium
(Fig.~\ref{rins}, solid line). Alternatively, this can be properly adjusted
so that the generated radiation may grow over a greater length, and the
conversion efficiency may become substantially larger as well. Consequently,
conversion efficiency of the input green radiation to VUV range, which is
greater than unity, becomes possible for difference-frequency process.

\begin{figure}[!h]
\begin{center}
\includegraphics[width=.4\columnwidth]{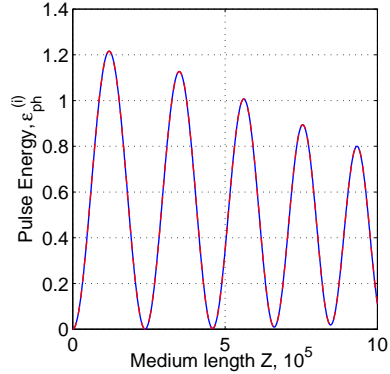}
\caption{\label{2gdownz} Same as in Fig.\ref{gdownz}(a), but
$|d_{ln}/d_{gl}|=1.1$ ($K_2=0.4$). }
\end{center}
\end{figure}

Simulations show that, besides other parameters, the actual power, energy
and photon conversion efficiencies are strongly dependent on the strengths
of the involved transitions and on their ratios (parameters $a$ and $K_j$).
Figure~\ref{2gdownz} displays the same results as in Fig.~\ref{gdownz}(a)
except that the ratio $|d_{ln}/d_{gl}|=0.35$ is changed for 1.1.
Consequently, the parameter $K_2$ changes from  0.04 to 0.4. The squared
modulus of the nonlinear polarization, which determines the FWM generated
power, is proportional to the product of squared modules of these transition
electric dipole matrix elements. Consequently, the output generated power
increases as well. (The squared modulus of the output Rabi frequency,
plotted in some of the graphs, increases by several orders even at the same
output power.) Therefore, the major features of the generation, optimum
conditions and and maximum output power depend on the specific quantum
system employed.

The characteristic absorption length
$z_0=(\alpha_{-}\tau_1/2\pi)^{-1}$ for Hg and the medium lengths
corresponding to Figs.~\ref{gdownz}-\ref{2gdownz} are estimated as
follows: The frequency-integrated absorption index is calculated as
$\alpha_-=\pi\Gamma_{gl}\alpha_{0-}=\lambda_s^2g_lA_{lg}N/4$, where
$\alpha_{0-}$ is the resonant value of the absorption index for the
monochromatic radiation, $\Gamma_{gl}$ is the half-width of the
resonance, $A_{lg}$ is the spontaneous relaxation rate at the
transition, $g_l$ is the degeneration factor for level $l$, and $N$
is the Hg number density. The oscillator strength for the Hg
transitions  $6s^2\,{}^1S_0 - 6p\,{}^1P_1$ ($\lambda=184.9$ nm) is
$f\approx 0.96$. Since $A_{lg}=0.67 g_g f_{lg}/g_{l}\lambda_{lg}^2,$
where $\lambda$ is given in cm, one obtains
$\alpha_{-}=(0.67g_gf_{lg}N/4)$ {cm}$^{-1}${c}$^{-1}$, and for
$\tau=3$ ns, and $N= 10^{16}$ cm$^{-3}$, an estimate gives
$\alpha_s\tau_1/2\pi\approx 7.6\times10^5 \hbox{ cm}^{-1}$.
Therefore, $Z=10^6$ corresponds to a medium length of about 1.3 cm.
\section{Conclusions}
We have presented an investigation of specific features of four-wave mixing
of delayed short pulses in two-photon resonant media, controlled by a
dynamic Stark shift of the resonance. Through numerical simulation, we have
analyzed different regimes of preparation of laser-induced maximum coherence
and its implementation for frequency conversion. We have shown that by a
proper choice of the time delays between the interacting laser pulses, the
conversion efficiencies may be enhanced by several orders of magnitude with
respect to conventional four-wave mixing. The results indicate the
feasibility of this technique to generate strong tunable vacuum-ultraviolet
radiation.
\section*{Acknowledgments} AKP, SAM and VVK  acknowledge support by EU INTAS
(project 99-00019) and by the Russian Foundation for Basic Research
(project~02-02-16325a). The authors are grateful to K. Bergmann and T.
Halfman for their statement of the problem and for valuable discussions over
the course of this work and on the manuscript. We  thank L. P. Yatsenko for
his comments on the manuscript of Ref. \cite{Yat01} prior to its
publication.


\end{document}